# Impacts of large-scale food fortification on the cost of nutrient-adequate diets: a modeling study in 89 countries

Leah Costlow*[1,2], Yan Bai*[3,4], Katherine P. Adams[5], Ty Beal[6], Kathryn G. Dewey[5], Christopher M. Free[7,8], Valerie M. Friesen[9] Mduduzi N. N. Mbuya[6], Stella Nordhagen[6], Florencia C. Vasta[6], William A. Masters[10,11]

**Affiliations**
* Contact authors: costlow1@msu.edu; ybai@worldbank.org
1. Dept. of Agricultural, Food, and Resource Economics, Michigan State Univ., East Lansing MI, USA
2. Dept. of Food Science and Human Nutrition, Michigan State Univ., East Lansing MI, USA
3. The World Bank, Development Data Group, Washington D.C., USA
4. Zhejiang University, School of Public Affairs, Hangzhou, Zhejiang, China
5. Institute for Global Nutrition, Dept. of Nutrition, Univ. of California Davis, Davis CA, USA
6. Global Alliance for Improved Nutrition, Washington, D.C., USA
7. Marine Science Institute, Univ. of California, Santa Barbara, Santa Barbara CA, 93106, USA
8. Bren School of Environmental Science and Management, Univ. of California, Santa Barbara, Santa Barbara CA, 93106, USA
9. Global Alliance for Improved Nutrition, London, UK
10. Friedman School of Nutrition Science and Policy, Tufts University, Boston MA, USA
11. Department of Economics, Tufts University, Medford MA, USA

**Acknowledgments**
This work was funded by the Swiss Agency for Development and Cooperation and the Gates Foundation, using methods developed through the Food Prices for Nutrition project at Tufts University supported by the Gates Foundation as INV-016158 (2000-25) and INV-053346 (2025-29). This work was done while the first author was a doctoral student in the Friedman School of Nutrition Science and Policy at Tufts University.

**Author contribution statement**
T.B., V.M.F., and W.A.M. conceived the study and obtained funding. L.C., Y.B., and W.A.M. developed the analytical approach. L.C. led the analysis and data visualization. L.C. and Y.B. drafted the paper. K.P.A., T.B., K.G.D., C.M.F., V.M.F., M.N.N.M., S.N., F.C.V., and W.A.M. provided critical inputs during the analysis process, gave feedback on the manuscript, and read and approved the final paper.

**Data and code availability**
All data analysis and visualizations were conducted in RStudio (version 2026.01.0 + 392). All R scripts (for main analysis, sensitivity analysis, and appendix materials) will be available at https://doi.org/10.17605/OSF.IO/TPRJH, as will nutrient requirements and derived modeled diet cost results. Retail food price data from the International Comparisons Program (ICP) coordinated by the World Bank are proprietary and not available for dissemination. For questions about the ICP food price data, readers may contact icp@worldbank.org. Fortification data will be made available upon request from the Global Fortification Data Exchange (GFDx).

# Impacts of large-scale food fortification on the cost of nutrient-adequate diets: a modeling study in 89 countries

## Abstract

**Introduction:** Large-scale food fortification (LSFF) is a widely accepted intervention to alleviate micronutrient deficiencies, yet policy implementation is often incomplete. Even assuming high implementation of existing mandatory and voluntary standards, the effects of LSFF on least-cost diets have not been established. Diet cost metrics such as the Cost of Nutrient Adequacy (CoNA) offer insights into how policy interventions such as LSFF might change access to nutritious diets around the world.

**Methods:** We estimated the extent to which LSFF reduces the retail cost of nutrient-adequate diets using retail food prices and fortification policy data from 89 countries. In total, we modeled 5,874 complete pairs (without fortification vs. with fortification) of least-cost diets across 22 sex-age groups and 3 nutrient-adequacy scenarios: meeting nutrient requirements only; adding minimum intakes for starchy staples and fruits and vegetables; and aligning food group shares with national consumption patterns. Single-vehicle and single-nutrient scenarios highlighted the cost-reductions associated with specific fortification approaches. In addition, we used regression methods to estimate the relationship between diet cost reductions and the total quantity of nutrients added to the food supply.

**Results:** Assuming 90% implementation of existing LSFF standards for wheat flour, maize flour, rice, and oil, we found median retail cost reductions of 1.6%, 2.4%, and 4.5% across the three scenarios. Cost reductions varied widely by sex-age groups, national fortification strategies and food price structures. Results from the regression analysis revealed that intermediate levels of food fortification resulted in the largest potential diet cost reductions.

**Conclusions:** These findings highlight that LSFF may improve economic access to nutritious diets when policies are carefully designed for local contexts, making it a valuable complement to other efforts that improve access to nutritious diets.

# Impacts of large-scale food fortification on the cost of nutrient-adequate diets: a modeling study in 89 countries

## Importance of the study

**What is already known on this topic**

- Large-scale food fortification (LSFF) is a proven tool for reducing the prevalence of micronutrient deficiency at relatively low cost, but food supplies are often not in full alignment with existing mandatory and voluntary fortification standards.
- Nutrient-adequate and healthy diets are too expensive for many low-income people around the world to purchase, which suggests that LSFF could help improve nutritional status without imposing new cost burdens.

**What this study adds**

- In the first study of its kind, we estimated the reduction in the modeled costs of nutrient-adequate diets that would result from 90% implementation of existing LSFF regulations, with a baseline comparison scenario assuming no fortification and models capturing the varying nutritional needs of different population subgroups.
- Fortification could reduce the cost of nutritious diets, but the size of the reduction depends on how fortification policies are specified and on how expensive nutritious diets were prior to fortification, which is linked to underlying food price dynamics.

**How this study might affect research, practice or policy**

- Fortification policies should be viewed as occupying a complementary role alongside broader food system or market-based efforts to expand economic access to healthy diets.

# Impacts of large-scale food fortification on the cost of nutrient-adequate diets: a modeling study in 89 countries

## Introduction

Micronutrient deficiencies affect billions of people worldwide[1]. They compromise immune and endocrine function, impair cognitive development, and increase the risk of maternal and child morbidity and mortality[2]. Despite advances in global health and nutrition, such deficiencies persist, particularly among vulnerable populations including young children and women of reproductive age[3]. Addressing micronutrient gaps at the population level is thus a global health imperative. Large-scale food fortification (LSFF), which involves adding essential vitamins and minerals to widely consumed staple foods, beverages, or condiments, has emerged as an effective strategy for reducing micronutrient deficiencies at the population level[4–8]. Yet while the health benefits of LSFF are well documented, its potential to lower the cost of nutrient-adequate diets and improve economic access to adequate nutrition has received limited attention.

Policies to standardize the fortification of commonly consumed staple foods or condiments have been widely instituted in countries across all income levels to ensure sufficient intake of essential nutrients that might otherwise be under-consumed in average diets[4,9,10]. However, biomarker data and dietary intake data show that micronutrient deficiencies remain prevalent[1,3], and in many countries, existing LSFF mandates have not been fully implemented, while in other cases fortification standards are voluntary and industry participation may be low[11]. When food supplies do not comply with fortification standards, nutrient sufficiency can only be achieved by consuming larger quantities of nutrient-rich foods, including fruits, vegetables, and animal-source foods, which tend to be more expensive per unit of energy than fortified or unfortified staple foods[12]. Healthy diets that are aligned with food-based dietary guidelines provide essential nutrients in the recommended amounts, but remain unaffordable for nearly 3 billion people around the world[13-15] and are more expensive than the cost of diets that meet nutrient requirements without necessarily providing sufficient quantities of all recommended food groups. This context highlights the importance of assessing whether LSFF can serve as a cost-effective strategy to increase the affordability of nutrient-adequate diets. Fortification could make low-cost staple foods more expensive if implementation costs are high and if high pass-through rates result in higher retail prices for consumers. Conversely, fortification could also succeed in providing low-cost sources of otherwise costly nutrients, resulting in least-cost diets that meet nutrient requirements at a lower overall cost.

Past research has established the health benefits of LSFF[7], and numerous modeling studies have investigated the cost-effectiveness or benefit-cost ratios of LSFF implementation in countries around the world.[16] However, there has not yet been a comprehensive study combining globally comparable price and fortification policy data and using linear optimization models to estimate the role of LSFF in improving economic access to nutritious diets. Our analysis addresses this gap by building on least-cost diet modeling methods that have previously been used to identify cases where key nutrients are unusually expensive to include in the diet, to identify differences in the cost of meeting nutrient needs across the life cycle, and to highlight seasonal variations in nutrient costs.[17–19] The base scenario in this analysis is the cost of a nutrient-adequate diet, whose value for a reference adult female has been monitored by the World Bank as a global indicator, with estimates now reported across countries to inform food and nutrition security monitoring efforts.[20] This study adapts these modeling tools to estimate the extent to which more complete implementation and enforcement of country-level LSFF policies would reduce the cost of nutrient-adequate diets and improve food and nutrition security around the world.

In this analysis, we calculated the cost of nutrient adequacy (CoNA) first assuming zero LSFF implementation and secondly assuming 90% implementation to estimate the reduction in diet costs due to fortification across three variations on a nutrient-adequate diet. The term "implementation" is used to describe compliance, i.e. the extent to which food processers comply with mandatory and voluntary fortification policies and produce foods that are fortified to specified levels. This provides insights into how implementation of fortification standards might improve access to nutrient-adequate diets and whether existing standards are optimally designed for this purpose. We estimated a total of 5,874 diet cost reductions for 89 countries and 22 population subgroups, iterating across three model specifications: (1) a basic model that identifies least-cost diets meeting nutrient requirements by subgroup; (2) an extension that additionally requires minimum intake of starchy staples (SS) and fruits and vegetables (FV); and (3) a further extension that sets minimum energy shares from major food groups reflecting consumption patterns in each country based on supply and utilization accounts (SUA) from FAO's Food Balance Sheets.[21] All models used the most recent 2021 global retail food prices from the International Comparison Program[22], coordinated by the World Bank, and LSFF policy data through 2022 from the Global Fortification Data Exchange (GFDx).[10]

## Materials and Methods

### Diet cost models

We calculated CoNA across three modeling scenarios using linear programming in a method adapted from previous publications[17,19]. This method minimizes the total cost of a food basket while ensuring compliance with a set of nutritional constraints, including targets for macronutrients, micronutrients, and total dietary energy. We updated the original CoNA method to include food group constraints in two additional scenarios. In all three scenarios, we included nutrient constraints for intakes of macronutrients and essential micronutrients (see Appendix Table A1 for a complete list). Nutrient requirements for 22 subgroups by age, sex, pregnancy status, and lactation status (hereafter referred to as "population subgroups") were from the Harmonized Nutrient Reference Values as defined in Allen et al.[23] and included average requirement and Tolerable Upper Intake Levels (ULs) for each nutrient as relevant. Diets meeting the needs of children under the age of four years were not included in this analysis. Energy requirements were calculated to align estimates from the European Food Safety Authority (EFSA)[24] with as defined in Allen et al. and assumed moderate physical activity (PAL = 1.8).

All three models included the same core constraints of nutrient adequacy. In the first scenario (CoNA), modeled diet costs were calculated to meet basic nutrient requirements with no additional food group requirements. This scenario represents the lower bound of nutrient adequacy by meeting nutrient needs without respect for food-based recommendations or cultural preferences. However, since basic nutrient adequacy does not result in modeled diets that broadly conform to culinary or cultural expectations about a balanced diet, including items from a range of particular food groups, we assessed two additional scenarios incorporating food group constraints to ensure that a given amount of energy from specified food groups was included in each modeled diet.

In the second scenario (CoNA-SS&FV), modeled diets were additionally required to meet a minimum threshold for energy from fruits and vegetables (combined, abbreviated as FV) and starchy staples (SS). Minimum intakes for these food groups were derived from the Healthy Diet Basket (HDB), which reflects commonalities across international food-based dietary guidelines and serves as the basis for global monitoring of the cost of a healthy diet[14]. The HDB provides reference intakes in kilocalories per day for a healthy, non-pregnant, non-lactating 30-year-old woman. We scaled target food group intakes up or down to accommodate changes in dietary energy requirements for other population subgroups. This scenario represents the cost to purchase nutrient-adequate diets while also satisfying the core nutrition-sensitive requirements of food-based dietary guidelines.

In the third scenario (CoNA-SUA), modeled diets instead were required to provide energy from four food groups recommended in the HDB, but within lower and upper bounds of national average consumption rather than in the quantities recommended for health. This scenario included food group constraints for fruits and vegetables (combined), starchy staples, legumes, nuts, and seeds (combined), and animal-source foods. Food supply quantities of energy from each food group were calculated at the country level from the FAO Supply Utilization Accounts (SUA)[21] and converted to energy shares using the daily energy requirements of each population subgroup. Upper and lower bounds for food group energy shares are calculated as the interquartile range (25$^{th}$ through 75$^{th}$ percentiles) of the food supply reported between 2010 and 2022. This scenario represents the cost to purchase nutrient-adequate diets that broadly conform to existing country-level consumption patterns by food group, assuming that food group energy shares are consistent across population subgroups.

For each of the three modeled diet scenarios, we first estimated diet costs assuming no fortification; then recalculated diet costs under the LSFF implementation scenario with increased nutrient density for fortified foods; then calculated the change in diet costs attributable to fortification. In each case, resulting modeled diet costs reflect standard matching of food composition data to non-fortified items. Foods that are potentially subject to fortification (“food vehicles”) include rice, oil, maize flour, and wheat flour, and were identified as fortifiable or not fortifiable at the country level based on the documentation of mandatory or voluntary fortification standards in GFDx. Retail food items were manually matched to food vehicles as appropriate based on item descriptions (see Appendix Table A2 for examples of the matching process).

Our modeling of existing fortification policies depends on several key assumptions related to 1) fortification policies and 2) premix costs. Although ready-made foods may include fortified ingredients, this varies widely by country and is not documented in GFDx. Accordingly, we assume that packaged, processed, and other ready-made foods with the exception of white bread are not subject to fortification, even if they may contain ingredients that are matchable to fortifiable food vehicles in a given country. For example, bagged wheat flour, white bread, and bottled vegetable oil are potentially fortifiable while whole grain bread, crackers, and sardines packed in oil may not. In most countries whole wheat flour is not subject to fortification based on GFDx documentation, with the exception of atta in India and Bangladesh. Fortified salt was excluded from this analysis and thus not allowed to enter into any of the modeled diet scenarios, along with all other non-caloric food items, spices, condiments, beverages, and foods that are not recommended as part of a healthy diet, such as sweets, snacks, and cured meats.

**Regression model**

We used a regression approach to describe how diet cost reductions are associated with baseline diet costs and with intensity of fortification policy, where policy intensity is estimated as the sum of all potentially fortifiable nutrients across all food vehicles in each country. This section of the analysis supplements the main diet cost modeling results and provides insights into whether countries that add larger quantities of nutrients to the food supply have larger potential cost reductions. A simple quadratic regression model reflects the likelihood that higher levels of fortification provide greater cost reductions up to an inflection point, after which the addition of further nutrients to the food supply is no longer efficient and results in higher costs due to the sensitivity of ULs on certain nutrients. All models included fixed effects for diet cost model and geographic region as shown in Equation 1:

$$\Delta C = \beta_0 + \beta_1 X_i + r_i + m_i + e_i \qquad (1)$$

where $\Delta C$ is the estimated percentage change in diet costs, $X_i$ is either the intensity of fortification or baseline diet costs in country $i$, and $r_i$ and $m_i$ are country-level fixed effects for World Bank region and diet cost model. Policy intensity was represented as the total combined amount of all fortified nutrients added across all food vehicles in each country, measured in mg per kg.

**Data**

We used retail food prices from 2021 from the International Comparisons Program (ICP), coordinated by the World Bank[22]. This data set includes 173 countries and was made available to the Food Prices for Nutrition project under a confidential agreement with the World Bank. Retail prices from ICP represent globally and regionally standardized item lists for cross-country comparisons. Information on mandatory and voluntary fortification policies is available for 154 countries from the Global Fortification Data Exchange (GFDx)[10], including data on the food vehicle, nutrient, and fortification amounts specified in LSFF policies in each country. National food supply data are available from FAO for 190 countries in total between 2010 and 2022. The final analytical sample of countries with data from all three data sources included 89 countries (see Appendix Table A4 for a complete list of countries in the analysis). Estimated reductions in diet costs are presented as global medians due to measurement error in country-level retail food prices and to account for notable outliers in the estimated reductions. Country inclusion and data on fortification coverage by food vehicle and micronutrient are summarized in the Appendix Table A5.

Implementation of existing LSFF policies varies widely across countries and within countries across food vehicles[10]. We modeled a hypothetical fortification scenario in which implementation of fortification mandates achieves 90% compliance with all documented policies, consistent with the improved compliance fortification scenarios modeled in Friesen et al (2026)[7]. Current implementation of fortification standards falls short of this 90% threshold on average, even for mandatory standards; while implementation approaches 100% for some mandatory standards in some places, typical levels of implementation range between 25% and 80%[7]. To calculate diet costs with LSFF implementation, we adjusted food composition for foods subject to fortification standards by adding 90% of the fortification amount to the baseline content of the fortified nutrient. In cases where multiple fortification standards are documented in GFDx for the same nutrient in a single country, we used the lowest reported level as a single decision rule, resulting in a conservative estimate of the lower-bound for potential fortification. In countries where wheat flour is fortified, we assumed that fortified wheat flour accounts for 75% of fortified white bread by weight, an estimate that is aligned with previous LSFF modeling studies[26]. The difference in cost between LSFF implementation and baseline scenarios represents the cost reduction attributable to LSFF. All costs were converted to 2021 international dollars using purchasing power parity (PPP) conversion factors[22]. We accounted for implementation costs by adding the cost of producing fortification nutrient premix to the ICP retail price for foods subject to fortification, assuming 100% pass-through of premix costs to retail prices which results in a higher-bound estimate of the effect of fortification on food prices. However, premix costs are low overall and account for a mean value of 0.1% of retail food prices across all food vehicles (see Appendix Figure A4). Premix costs were estimated at the country and food vehicle levels and account for micronutrient type, fortification amount, chemical compound (fortificant), and supply chain costs as reported in Friesen et al (2026).[7] We did not estimate the additional implementation and administrative costs that the food industry or national governments would likely incur. See Appendix Table A6 for results of a sensitivity analysis using alternative pass-through (50%) and LSFF implementation (50%) rates.

**Results**

Implementation of existing country-level fortification policies resulted in reduced diet costs on average across all subgroups included in the analysis (see Table 1 for average regional cost results). Diet cost estimates were successfully obtained for nearly all cases, but solutions were infeasible in 0.6% of all unique country-subgroup-scenario-fortification combinations (37 out of 11,748; see Appendix Table A7 for a list of cases with no solution). Without fortification, CoNA had the lowest median daily cost at $2.70 (interquartile range [IQR]: $2.20 to $3.23) across all countries and population subgroups. The addition of two food group constraints in the CoNA-SS&FV scenario raised the median daily cost to $3.09 (IQR: $2.52 to $3.83). The CoNA-SUA scenario had the highest median daily cost at $3.60 (IQR: $2.97 to $4.52). In comparison, the Cost of a Healthy Diet (CoHD) meeting food-based dietary guidelines was a median of $3.48 (IQR: $3.05 to $3.88) according to official 2021 estimates.[20] The relatively higher cost of the CoNA-SUA scenario reflects wide variation in observed food group consumption shares, as well as potential selection effects due to the smaller sample size of the present analysis.

Implementation of LSFF at 90% of existing standards reduced diet costs for CoNA by an average across all countries and population subgroups of $0.04 or about 1.6% per day (IQR: -2.4% to 0%), for CoNA-SS&FV by $0.08 or about 2.4% per day (IQR: -4.3% to 0%), and for CoNA-SUA by $0.16 or about 4.5% per day (IQR -8.8% to -0.1%). In Figure 1, Panel A depicts the range of changes in the cost of a nutrient-adequate diet across all 89 countries in the analytical sample, while Panel B depicts the range of cost changes among only the subset of countries with a non-zero change in diet costs under LSFF implementation (see within-figure labels for sample size by scenario). Within the 5,874 modeled diet pairs across 89 countries, 22 population subgroups, and 3 scenarios, fortification reduced diet costs in 54.8% of modeled

diets, had no effect in 38.0%, and led to higher diet costs in 7.3%, while no feasible solution was obtained in 0.6% (37 cases). Within the full analytical sample, 90% implementation of fortification standards was responsible for a very small median cost reduction across all population subgroups and modeled diet scenarios of 0.1% (IQR: -4.6% to 0%). Among only those countries and scenarios with non-zero cost changes, median reductions were larger at 2.4% (IQR: -9.0% to -0.3%).

Implementation of LSFF resulted in larger diet cost reductions for certain population subgroups compared to others (Figure 1). For countries where implementing LSFF results in a change in diet costs, median reductions were slightly larger for female subgroups at 2.6% (IQR: -9.5% to -0.3%) compared to 2.1% (IQR: -8.1% to -0.2%) for male subgroups. Within female subgroups, median cost reductions were smallest during lactation at 1.7% (IQR: -7.3% to -0.2%), larger during pregnancy at 2.2% (IQR: -8.4% to -0.3%), and largest for females of all ages who are not pregnant or lactating at 3.2% (IQR: -10.7% to -0.3%). The range of reductions was widest for females of all ages, boys aged 4-6 years, adolescent boys aged 11-14 years, and men aged 71-79 years.

Cost reductions also varied across the three model specification scenarios, although there was substantial overlap. Under the nutrients-only CoNA model, median cost reductions ranged from 0.5% for girls 18 years old and younger during lactation and pregnancy to 3.4% for men aged 71-79 years. Under the CoNA-SS&FV model, median reductions ranged from 0.7% for boys aged 15 to 17 years to 5.1% for women aged 25 to 50 years. The largest reductions occurred under the CoNA-SUA model, with median reductions ranging from 2.6% for girls 18 years old and younger during lactation to 7.3% for girls aged 11-14 years.

Median cost changes at the subgroup level were uniformly negative (i.e., LSFF implementation consistently resulted in reduced rather than increased diet costs). Positive values located outside of the interquartile region of the boxplots (i.e. on the whiskers) in Figure 1 reflect country-subgroup combinations where fortification resulted in higher diet costs. This may result from two main causes in different cases: first, in some cases fortification of nutrients with ULs on their recommended consumption may lead modeled diets to include quantities of more expensive, non-fortified foods in order to meet minimum requirements for other nutrients while remaining below the UL for the fortified nutrient. Underlying patterns in country-level retail prices also contribute, as there may not be a substantial increase in diet costs if the alternative item is similarly low-cost.

**Figure 1. Change in cost of nutrient-adequate diets with LSFF implementation by scenario and population subgroups.**

**Panel A. All countries**

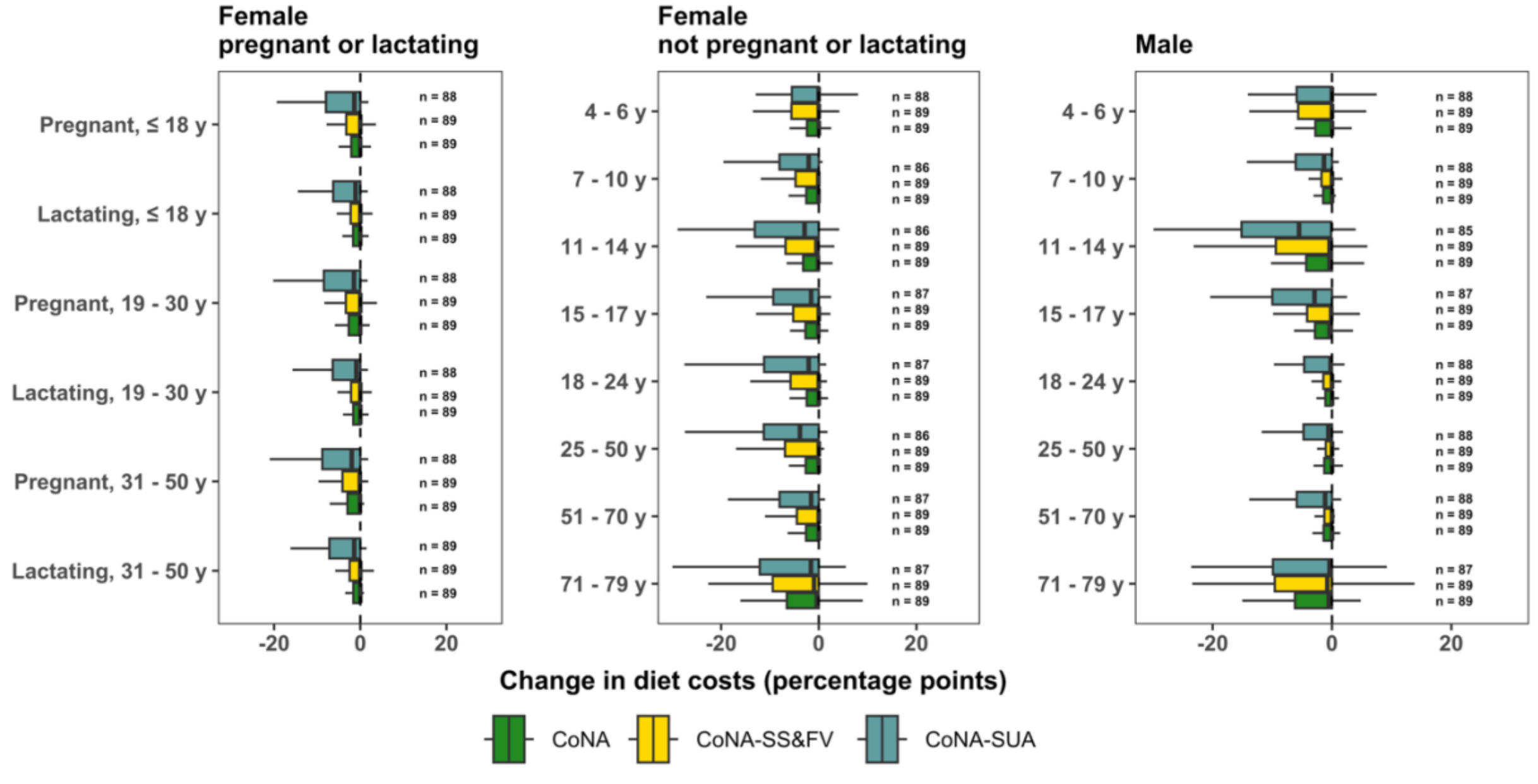

**Panel B. Countries with any change in diet costs**

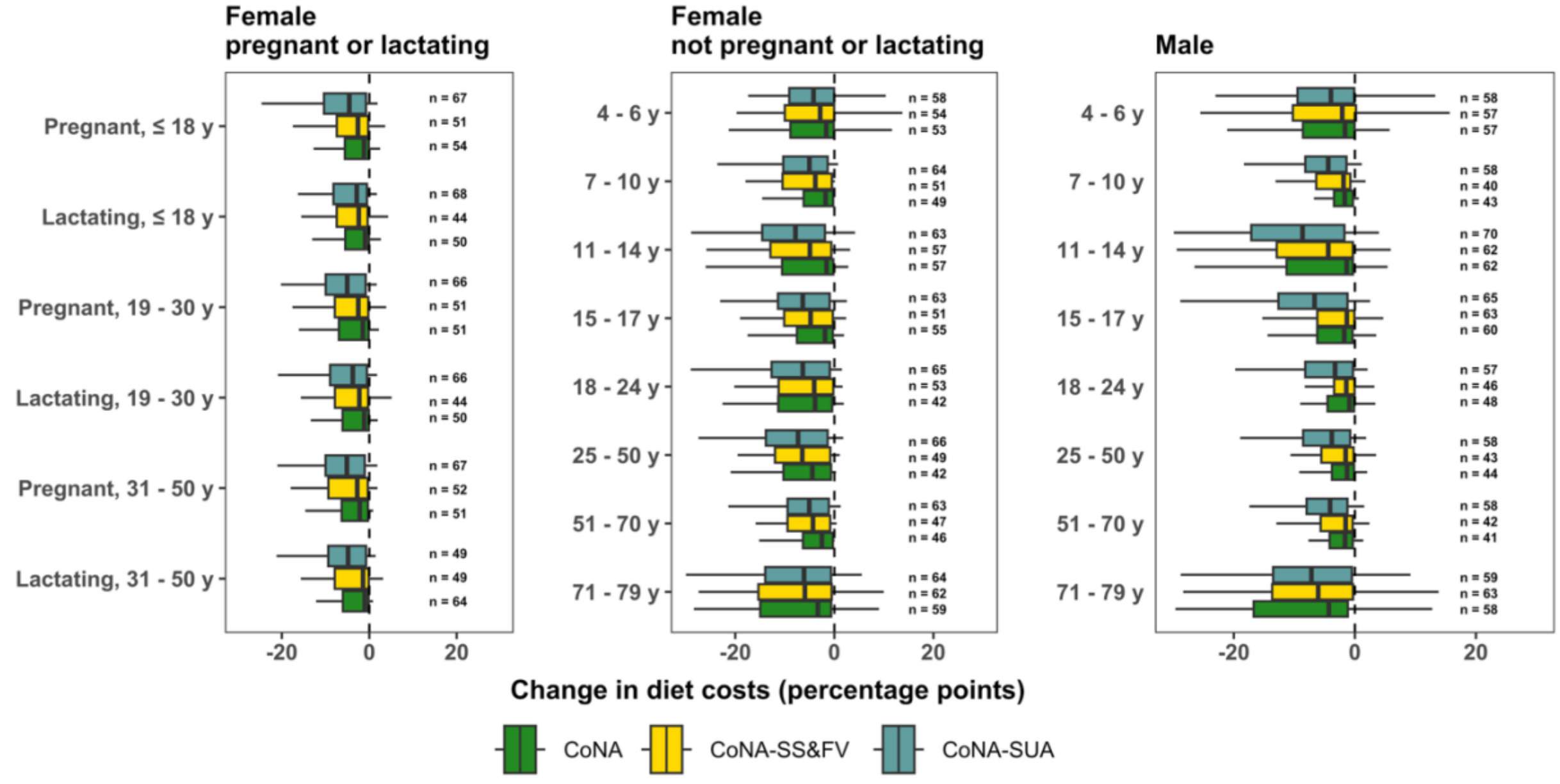

**Note:** Data shown in Panel A are changes in diet costs from all countries in the analytical dataset with non-missing solutions to all modeled diets, while data shown in Panel B are changes in diet costs from a varying subset of countries with non-zero change in diet costs under LSFF implementation. Negative values indicate reduced costs with LSFF while positive values indicate increased costs with LSFF. CoNA refers to diets meeting nutrient requirements; CoNA-SS&FV refers to diets that also meet basic healthy food group constraints for starchy staples, fruits, and vegetables; and CoNA-SUA refers to diets that align with national average consumption patterns at the food group level. Outliers are omitted from all boxplots using the Tukey standard (1.5 x IQR), as are 37 cases with no feasible solution.

Across all 3 scenarios, reduced diet costs resulted from changes to the composition of the least-cost food basket (Figure 2). Substitution between fortified and unfortified versions of the same items was the primary driver of reduced diet costs, with only minimal additional changes in the composition of modeled diet baskets. The increased nutrient density of fortified foods means that these items were included in large quantities, while the inclusion of unfortified staple foods decreased. In cases where non-fortified item inclusion increased with fortification, this indicates that the item did not contribute nutrients at minimum cost in the baseline CoNA scenario, and that the introduction of fortification provided greater flexibility in the linear programming algorithm for different foods to be included. For example, broad beans, spotted beans, mangoes, chicken, and milk are all included in fortified diet baskets to a larger degree than in unfortified diet baskets. The inclusion of other legumes, vegetables, and animal-source foods declined to a degree that largely offsets increases elsewhere such that, on average, food group energy remained roughly constant both with and without fortification (see Appendix Figure A1). This is possible because the modeled diets include specified food groups in amounts that are at least equal to benchmarks, allowing for variation in food group energy across scenarios.

**Figure 2. Changes in modeled diet composition by item with LSFF implementation by scenario and population subgroup (percentage of recommended energy intake by food group).**

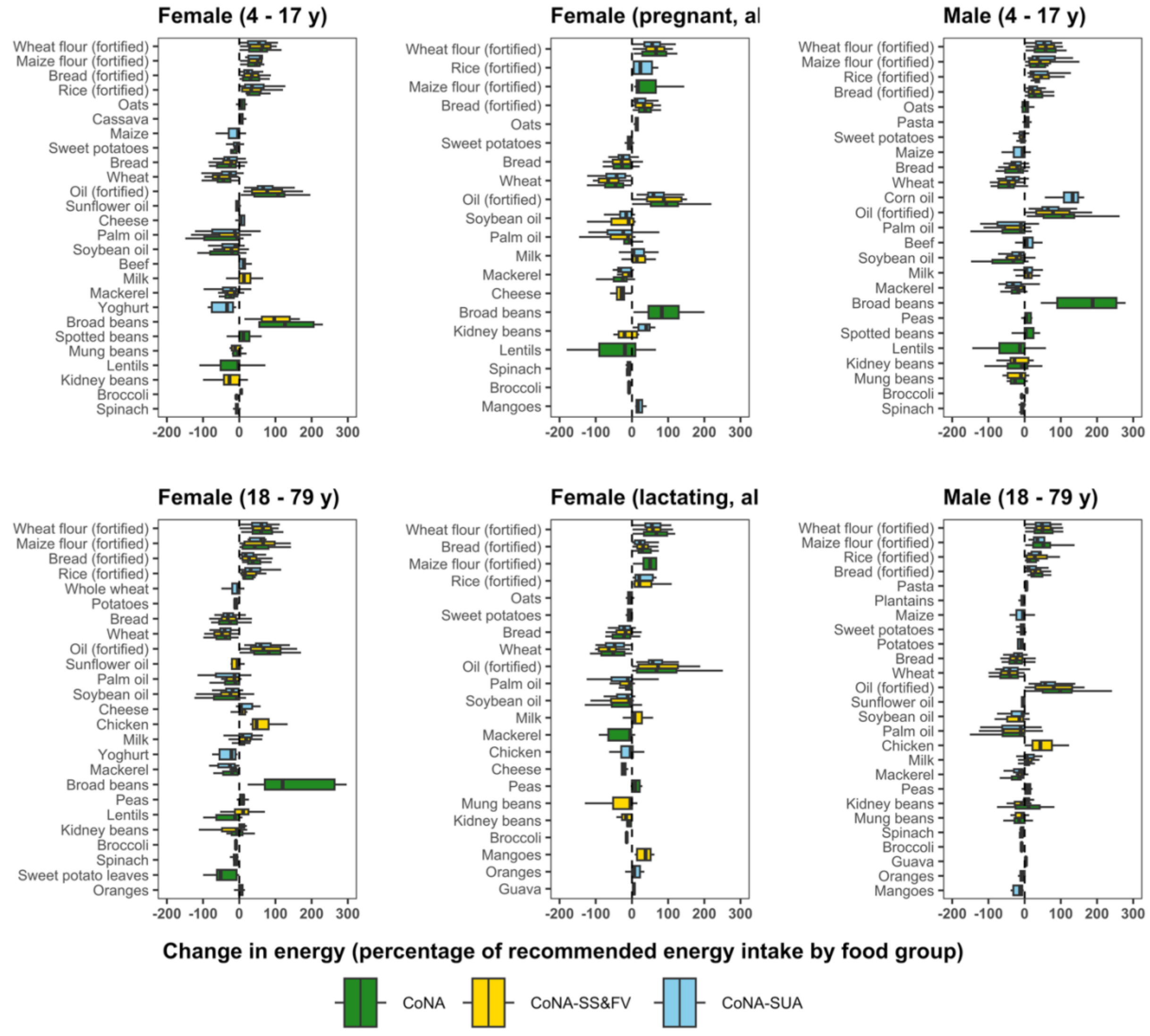

**Note:** Data shown are changes in average energy from items included in least-cost diets as a percentage of recommended energy intake by food group for each population subgroup, while population subgroups are combined for ease of presentation. Negative values indicate reduced costs with LSFF while positive values indicate increased costs with LSFF. Recommended intakes are derived from the Healthy Diet Basket and are scaled up or down for each subgroup's energy needs. Items whose median energy included changes by 3% or less under fortification are excluded from the visualization. CoNA refers to diets meeting nutrient requirements; CoNA-SS&FV refers to diets that also meet basic healthy food group constraints for starchy staples, fruits, and vegetables; and CoNA-SUA refers to diets that align with national average consumption patterns at the food group level. Outliers are omitted from all boxplots using the Tukey standard (1.5 x IQR).

Diet cost reductions were primarily driven by certain food vehicles more than others (Figure 3). These results reflect the isolated effect on diet costs of fortifying only one food vehicle at a time. Fortification of wheat flour, which was the most commonly fortified food vehicle (see Appendix Table A5), contributed the largest reduction in diet costs across all three scenarios and for all population subgroups. Rice fortification contributed to a lesser extent but still consistently led to diet cost reductions for most subgroups, especially in the CoNA-SUA scenario. On average, oil fortification had no effect on diet costs across all scenarios and subgroups, while maize flour fortification had no effect in most cases and led to higher diet costs for boys aged 4-6 years in the CoNA-SS&FV scenarios.

**Figure 3. Median percentage change in cost of nutrient-adequate diets with LSFF implementation by food vehicle and population subgroup.**

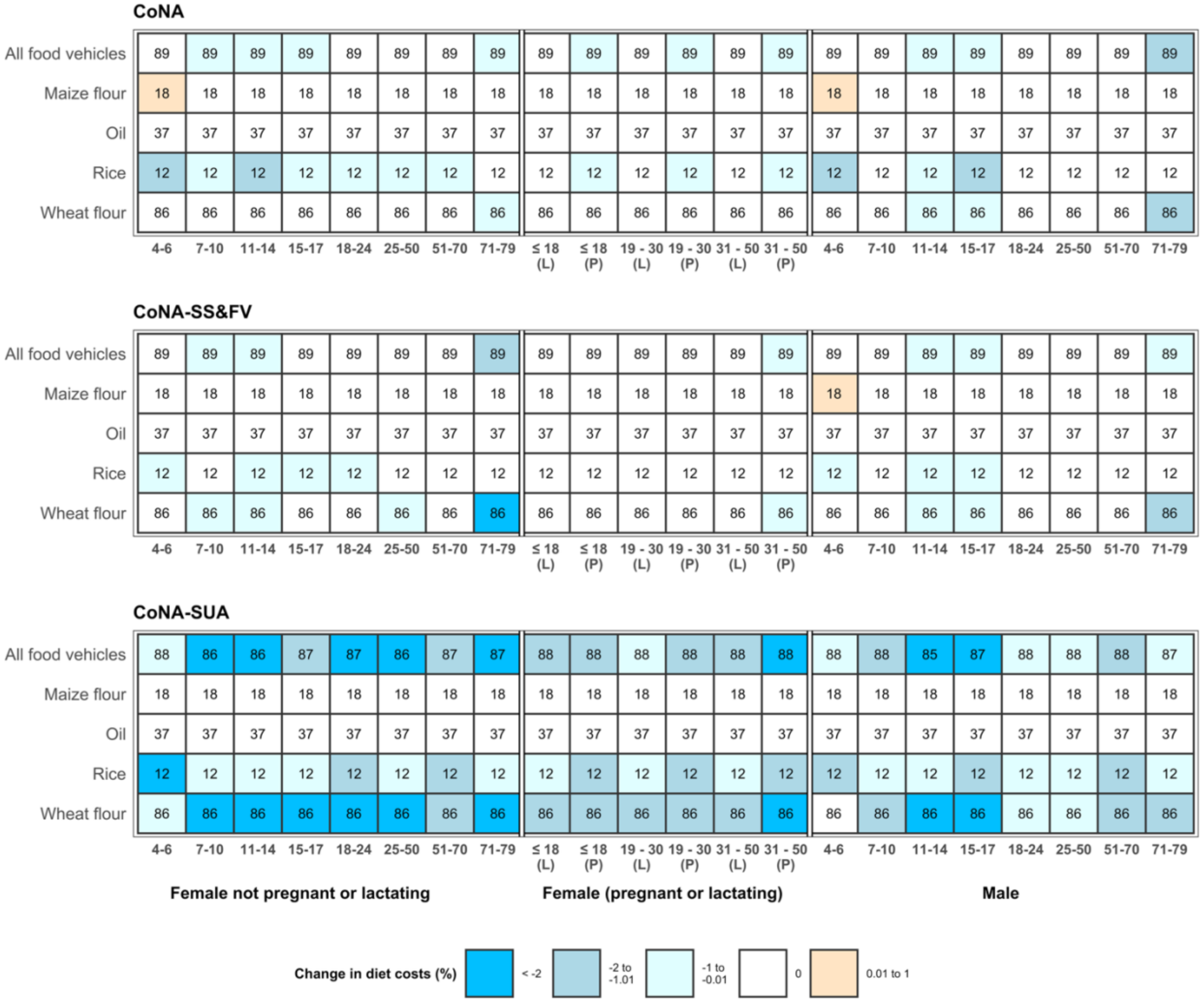

**Note:** Data shown are median reductions in diet cost by scenario, food vehicle, and population subgroup. Labels indicate sample size (n = countries). Numbers on the y-axis indicate age ranges in years. Cost reductions are calculated as percentage change in 2021 international dollars (PPP). CoNA refers to diets meeting nutrient requirements; CoNA-SS&FV refers to diets that also meet basic healthy food group constraints for starchy staples, fruits, and vegetables; and CoNA-SUA refers to diets that align with national average consumption patterns at the food group level. Cases with no feasible solution are omitted.

**Figure 4. Median percentage change in cost of nutrient-adequate diets with LSFF implementation by nutrient and population subgroup.**

**CoNA**

| | Female (not pregnant or lactating) | | | | | | | | Female (pregnant or lactating) | | | | | | Male | | | | | | | |
|---|---|---|---|---|---|---|---|---|---|---|---|---|---|---|---|---|---|---|---|---|---|---|
| | 4-6 | 7-10 | 11-14 | 15-17 | 18-24 | 25-50 | 51-70 | 71-79 | ≤ 18 (L) | ≤ 18 (P) | 19 - 30 (L) | 19 - 30 (P) | 31 - 50 (L) | 31 - 50 (P) | 4-6 | 7-10 | 11-14 | 15-17 | 18-24 | 25-50 | 51-70 | 71-79 |
| All nutrients | 89 | 89 | 89 | 89 | 89 | 89 | 89 | 89 | 89 | 89 | 89 | 89 | 89 | 89 | 89 | 89 | 89 | 89 | 89 | 89 | 89 | 89 |
| Calcium | 23 | 23 | 23 | 23 | 23 | 23 | 23 | 23 | 23 | 23 | 23 | 23 | 23 | 23 | 23 | 23 | 23 | 23 | 23 | 23 | 23 | 23 |
| Folate | 71 | 71 | 71 | 71 | 71 | 71 | 71 | 71 | 71 | 71 | 71 | 71 | 71 | 71 | 71 | 71 | 71 | 71 | 71 | 71 | 71 | 71 |
| Iron | 83 | 83 | 83 | 83 | 83 | 83 | 83 | 83 | 83 | 83 | 83 | 83 | 83 | 83 | 83 | 83 | 83 | 83 | 83 | 83 | 83 | 83 |
| Niacin | 61 | 61 | 61 | 61 | 61 | 61 | 61 | 61 | 61 | 61 | 61 | 61 | 61 | 61 | 61 | 61 | 61 | 61 | 61 | 61 | 61 | 61 |
| Riboflavin | 61 | 61 | 61 | 61 | 61 | 61 | 61 | 61 | 61 | 61 | 61 | 61 | 61 | 61 | 61 | 61 | 61 | 61 | 61 | 61 | 61 | 61 |
| Selenium | 2 | 2 | 2 | 2 | 2 | 2 | 2 | 2 | 2 | 2 | 2 | 2 | 2 | 2 | 2 | 2 | 2 | 2 | 2 | 2 | 2 | 2 |
| Thiamin | 63 | 63 | 63 | 63 | 63 | 63 | 63 | 63 | 63 | 63 | 63 | 63 | 63 | 63 | 63 | 63 | 63 | 63 | 63 | 63 | 63 | 63 |
| Vitamin A | 43 | 43 | 43 | 43 | 43 | 43 | 43 | 43 | 43 | 43 | 43 | 43 | 43 | 43 | 43 | 43 | 43 | 43 | 43 | 43 | 43 | 43 |
| Vitamin B12 | 25 | 25 | 25 | 25 | 25 | 25 | 25 | 25 | 25 | 25 | 25 | 25 | 25 | 25 | 25 | 25 | 25 | 25 | 25 | 25 | 25 | 25 |
| Vitamin B6 | 20 | 20 | 20 | 20 | 20 | 20 | 20 | 20 | 20 | 20 | 20 | 20 | 20 | 20 | 20 | 20 | 20 | 20 | 20 | 20 | 20 | 20 |
| Vitamin E | 3 | 3 | 3 | 3 | 3 | 3 | 3 | 3 | 3 | 3 | 3 | 3 | 3 | 3 | 3 | 3 | 3 | 3 | 3 | 3 | 3 | 3 |
| Zinc | 34 | 34 | 34 | 34 | 34 | 34 | 34 | 34 | 34 | 34 | 34 | 34 | 34 | 34 | 34 | 34 | 34 | 34 | 34 | 34 | 34 | 34 |

**CoNA-SS&FV**

| | Female (not pregnant or lactating) | | | | | | | | Female (pregnant or lactating) | | | | | | Male | | | | | | | |
|---|---|---|---|---|---|---|---|---|---|---|---|---|---|---|---|---|---|---|---|---|---|---|
| | 4-6 | 7-10 | 11-14 | 15-17 | 18-24 | 25-50 | 51-70 | 71-79 | ≤ 18 (L) | ≤ 18 (P) | 19 - 30 (L) | 19 - 30 (P) | 31 - 50 (L) | 31 - 50 (P) | 4-6 | 7-10 | 11-14 | 15-17 | 18-24 | 25-50 | 51-70 | 71-79 |
| All nutrients | 89 | 89 | 89 | 89 | 89 | 89 | 89 | 89 | 89 | 89 | 89 | 89 | 89 | 89 | 89 | 89 | 89 | 89 | 89 | 89 | 89 | 89 |
| Calcium | 23 | 23 | 23 | 23 | 23 | 23 | 23 | 23 | 23 | 23 | 23 | 23 | 23 | 23 | 23 | 23 | 23 | 23 | 23 | 23 | 23 | 23 |
| Folate | 71 | 71 | 71 | 71 | 71 | 71 | 71 | 71 | 71 | 71 | 71 | 71 | 71 | 71 | 71 | 71 | 71 | 71 | 71 | 71 | 71 | 71 |
| Iron | 83 | 83 | 83 | 83 | 83 | 83 | 83 | 83 | 83 | 83 | 83 | 83 | 83 | 83 | 83 | 83 | 83 | 83 | 83 | 83 | 83 | 83 |
| Niacin | 61 | 61 | 61 | 61 | 61 | 61 | 61 | 61 | 61 | 61 | 61 | 61 | 61 | 61 | 61 | 61 | 61 | 61 | 61 | 61 | 61 | 61 |
| Riboflavin | 61 | 61 | 61 | 61 | 61 | 61 | 61 | 61 | 61 | 61 | 61 | 61 | 61 | 61 | 61 | 61 | 61 | 61 | 61 | 61 | 61 | 61 |
| Selenium | 2 | 2 | 2 | 2 | 2 | 2 | 2 | 2 | 2 | 2 | 2 | 2 | 2 | 2 | 2 | 2 | 2 | 2 | 2 | 2 | 2 | 2 |
| Thiamin | 63 | 63 | 63 | 63 | 63 | 63 | 63 | 63 | 63 | 63 | 63 | 63 | 63 | 63 | 63 | 63 | 63 | 63 | 63 | 63 | 63 | 63 |
| Vitamin A | 43 | 43 | 43 | 43 | 43 | 43 | 43 | 43 | 43 | 43 | 43 | 43 | 43 | 43 | 43 | 43 | 43 | 43 | 43 | 43 | 43 | 43 |
| Vitamin B12 | 25 | 25 | 25 | 25 | 25 | 25 | 25 | 25 | 25 | 25 | 25 | 25 | 25 | 25 | 25 | 25 | 25 | 25 | 25 | 25 | 25 | 25 |
| Vitamin B6 | 20 | 20 | 20 | 20 | 20 | 20 | 20 | 20 | 20 | 20 | 20 | 20 | 20 | 20 | 20 | 20 | 20 | 20 | 20 | 20 | 20 | 20 |
| Vitamin E | 3 | 3 | 3 | 3 | 3 | 3 | 3 | 3 | 3 | 3 | 3 | 3 | 3 | 3 | 3 | 3 | 3 | 3 | 3 | 3 | 3 | 3 |
| Zinc | 34 | 34 | 34 | 34 | 34 | 34 | 34 | 34 | 34 | 34 | 34 | 34 | 34 | 34 | 34 | 34 | 34 | 34 | 34 | 34 | 34 | 34 |

**CoNA-SUA**

| | Female (not pregnant or lactating) | | | | | | | | Female (pregnant or lactating) | | | | | | Male | | | | | | | |
|---|---|---|---|---|---|---|---|---|---|---|---|---|---|---|---|---|---|---|---|---|---|---|
| | 4-6 | 7-10 | 11-14 | 15-17 | 18-24 | 25-50 | 51-70 | 71-79 | ≤ 18 (L) | ≤ 18 (P) | 19 - 30 (L) | 19 - 30 (P) | 31 - 50 (L) | 31 - 50 (P) | 4-6 | 7-10 | 11-14 | 15-17 | 18-24 | 25-50 | 51-70 | 71-79 |
| All nutrients | 88 | 86 | 86 | 87 | 87 | 86 | 87 | 87 | 88 | 88 | 88 | 88 | 88 | 88 | 88 | 88 | 85 | 87 | 88 | 88 | 88 | 87 |
| Calcium | 23 | 23 | 23 | 23 | 23 | 23 | 23 | 23 | 23 | 23 | 23 | 23 | 23 | 23 | 23 | 23 | 23 | 23 | 23 | 23 | 23 | 23 |
| Folate | 71 | 71 | 71 | 71 | 71 | 71 | 71 | 71 | 71 | 71 | 71 | 71 | 71 | 71 | 71 | 71 | 71 | 71 | 71 | 71 | 71 | 71 |
| Iron | 83 | 83 | 83 | 83 | 83 | 83 | 83 | 83 | 83 | 83 | 83 | 83 | 83 | 83 | 83 | 83 | 83 | 83 | 83 | 83 | 83 | 83 |
| Niacin | 61 | 61 | 61 | 61 | 61 | 61 | 61 | 61 | 61 | 61 | 61 | 61 | 61 | 61 | 61 | 61 | 61 | 61 | 61 | 61 | 61 | 61 |
| Riboflavin | 61 | 61 | 61 | 61 | 61 | 61 | 61 | 61 | 61 | 61 | 61 | 61 | 61 | 61 | 61 | 61 | 61 | 61 | 61 | 61 | 61 | 61 |
| Selenium | 2 | 2 | 2 | 2 | 2 | 2 | 2 | 2 | 2 | 2 | 2 | 2 | 2 | 2 | 2 | 2 | 2 | 2 | 2 | 2 | 2 | 2 |
| Thiamin | 63 | 63 | 63 | 63 | 63 | 63 | 63 | 63 | 63 | 63 | 63 | 63 | 63 | 63 | 63 | 63 | 63 | 63 | 63 | 63 | 63 | 63 |
| Vitamin A | 43 | 43 | 43 | 43 | 43 | 43 | 43 | 43 | 43 | 43 | 43 | 43 | 43 | 43 | 43 | 43 | 43 | 43 | 43 | 43 | 43 | 43 |
| Vitamin B12 | 25 | 25 | 25 | 25 | 25 | 25 | 25 | 25 | 25 | 25 | 25 | 25 | 25 | 25 | 25 | 25 | 25 | 25 | 25 | 25 | 25 | 25 |
| Vitamin B6 | 20 | 20 | 20 | 20 | 20 | 20 | 20 | 20 | 20 | 20 | 20 | 20 | 20 | 20 | 20 | 20 | 20 | 20 | 20 | 20 | 20 | 20 |
| Vitamin E | 3 | 3 | 3 | 3 | 3 | 3 | 3 | 3 | 3 | 3 | 3 | 3 | 3 | 3 | 3 | 3 | 3 | 3 | 3 | 3 | 3 | 3 |
| Zinc | 34 | 34 | 34 | 34 | 34 | 34 | 34 | 34 | 34 | 34 | 34 | 34 | 34 | 34 | 34 | 34 | 34 | 34 | 34 | 34 | 34 | 34 |

Change in diet costs (%): < -2 | -2 to -1.01 | -1 to -0.01 | 0 | 0.01 to 1 | > 1.01

**Note**: Data shown are median reductions in diet cost by scenario, nutrient, and sex-ag subgroup. Labels indicate sample size (n = number of countries). Numbers on the y-axis indicate age ranges in years. Cost reductions are calculated as percentage change in 2021 international dollars (PPP). CoNA refers to diets meeting nutrient requirements only; CoNA-SS&FV refers to diets that also meet healthy food group minimum intakes for starchy staples, fruits, and vegetables; and CoNA-SUA refers to diets that align with national average consumption patterns at the food group level. Cases with no feasible solution are omitted.

Changes in diet costs under LSFF implementation were also driven by certain nutrients more than others, while other nutrients contributed to increased diet costs (Figure 4). As in Figure 3, these results reflected the isolated effect on diet costs of fortifying one nutrient at a time. Fortification with calcium drove cost reductions in the CoNA and CoNA-SS&FV scenarios, with some contribution from vitamin E. Fortification with iron contributed to the slightly larger diet cost reductions in the CoNA-SUA scenario, again with some contribution from vitamin E. Fortification with selenium led to diet cost increases across all three scenarios, although only four countries had selenium fortification policies. Zinc fortification caused sporadic diet cost increases as well. For the remaining nutrients, median reductions were zero, indicating that single-nutrient fortification had no effect on least-cost diets.

Diet cost reductions across all three model scenarios have a nonlinear relationship with the intensity of a country's fortification policies and a largely negative association with baseline diet costs, as shown in Figure 5. Policy intensity is estimated as the sum of all potentially fortifiable nutrients across all food vehicles in each country. Countries with the largest reductions in diet costs tend to have intermediate combined fortification amounts (left-hand column), whereas those with the most extensive combined fortification amounts are more likely to have no cost reductions or even diet cost increases. Similarly, reductions are often greatest in countries with relatively high baseline diet costs prior to LSFF implementation (right-hand column). However, these correlations are not strictly consistent, as a wide range of diet cost reductions is observed across the distribution. In addition, diet cost reductions are further linked with geographic region. Countries in Latin America & Caribbean tend to have both higher diet cost reductions, intermediate fortification amounts, and relatively higher baseline diet costs. Countries in Sub-Saharan Africa tend to have lower or no diet cost reduction, higher fortification amounts, and more moderate baseline diet costs. Fortification leads to no change in diet costs for most countries in East Asia & Pacific or in Europe & Central Asia, although fewer European countries are included in the analytical dataset. Across all population subgroups, average cost reductions are largest for high income and upper middle income countries (Appendix Figure A5). However, due to the small sample size of the analysis and selection effects resulting in a low representation of high-income countries, these region- and income-level associations should be viewed cautiously. Underlying food price patterns may also be responsible for patterns in diet cost reductions; for example, if nutritious foods are already inexpensive in a given country, then LSFF implementation might not result in meaningfully lower modeled diet costs.

Several countries consistently exhibited the largest average reductions across all scenarios (Appendix Figure A2). For example, in the basic CoNA scenario, LSFF reduces median diet costs most significantly in Qatar (mean = 24.1%), Paraguay (20.7%), and Trinidad and Tobago (15.9%). In the CoNA-SS&FV scenario, LSFF reduces median diet costs most significantly in Paraguay (18.0%), Qatar (17.3%), and Uruguay (15.0%). Finally, in the CoNA-SUA scenario, LSFF leads to large median reductions in diet costs of greater than 25% in Colombia (34.3%), Uruguay (27.5%), and Argentina (26.9%), while a further 17 countries have reductions between 10% and 25%. The same countries tend to experience relatively substantial reductions under all three scenarios, indicating that fortification would be especially effective in increasing economic access to nutrient-adequate diets in these countries. In cases where fortification leads to diet cost increases for any given model and population subgroup, this may reflect inefficiently high fortification levels as implied in Figure 5, especially for younger children whose reference values for nutrients like zinc are relatively low (see Appendix Figure A3 for examples).

**Figure 5. Diet cost reductions compared with combined fortification quantities and baseline modeled diet costs by country and region.**

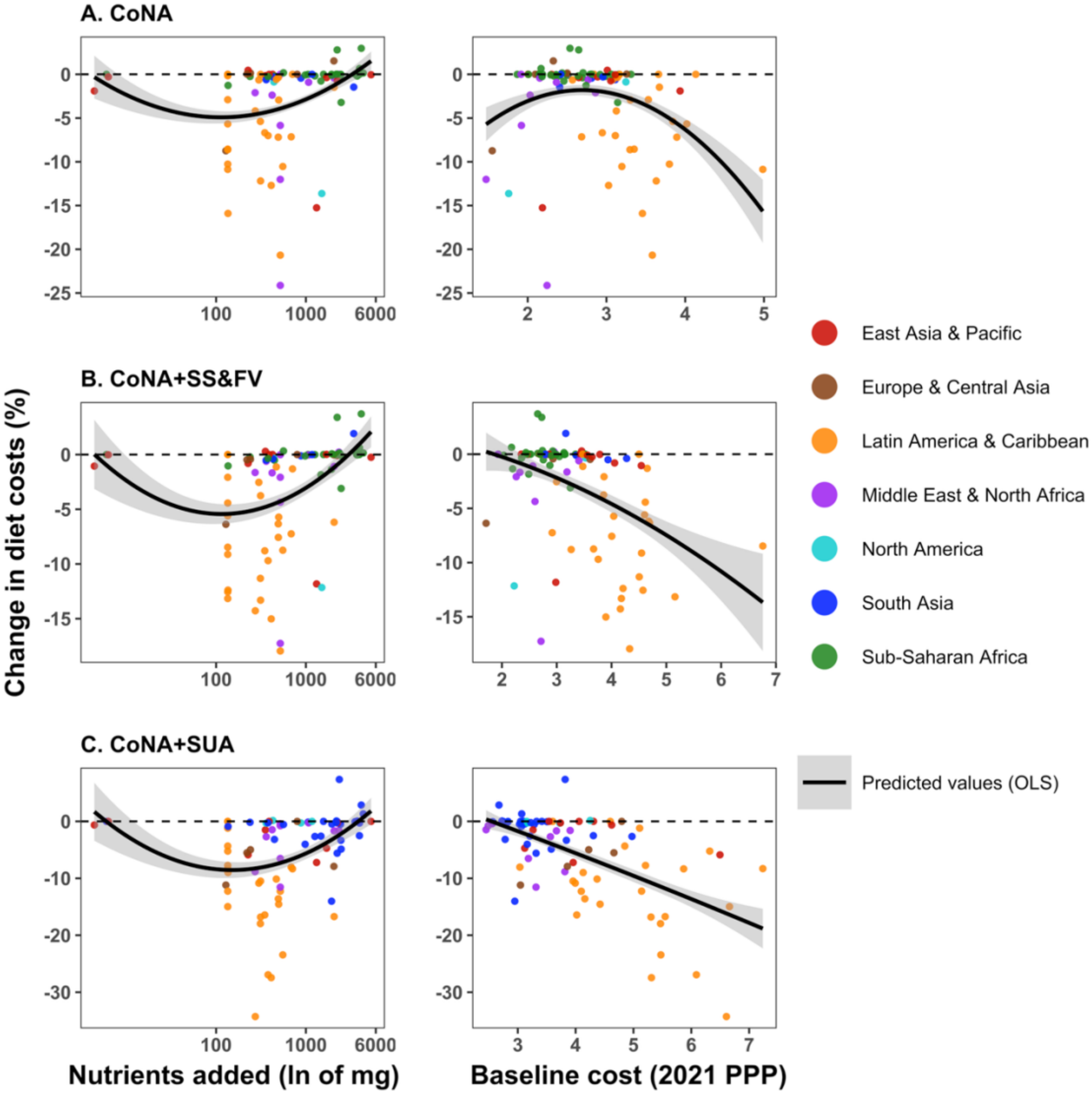

**Note:** Scatter plot values shown on the y-axis are median changes in country-level diet costs across all population subgroups, including each modeled diet scenario separately. Regions are from the World Bank Development Indicators. Black lines indicate predicted values from quadratic OLS regressions with fixed effects for region and diet cost model, and grey shaded areas indicate 95% confidence intervals. In the left-hand column, data shown on the x-axis are country-level totals in fortification amounts across all nutrients and food vehicles, expressed in mg of nutrient added per kg of food vehicle and shown on a log scale. In the right-hand column, data shown on the x-axis are country-level diet costs assuming no LSFF implementation. Cases with no feasible solution are omitted.

## Discussion

Our results reveal that implementation of existing LSFF standards leads to modest reductions in the cost of nutrient-adequate diets, with substantial variation across countries and sex-age groups, and with within-country differences further driven by specific food vehicles and nutrients more than others. The largest cost reductions are observed in diets that provide food group amounts that match current consumption patterns. In slightly more than one third of modeled diets, LSFF implementation had no impact on diet costs, while diet costs rise in a smaller number of modeled diets. Differences in the direction and magnitude of change reflect variation in national food prices and fortification policies, including the number of food vehicles fortified, range of nutrients mandated, fortification levels, and the dietary assumptions embedded in the least-cost modeling.

Single-vehicle and single-nutrient analyses reveal that, in select cases, fortification of a single food vehicle can effectively reduce nutrient-adequate diet costs, while fortification with a single nutrient often had little or no effect, with exceptions for calcium and iron. Among food vehicles, wheat flour fortification was the most effective in reducing diet costs. Among nutrients, iron and calcium contributed the most to cost reductions, indicating that these nutrients were the most costly to obtain even from least-cost sources within the optimized nutrient-adequate diet. The addition of other single nutrients to the food supply may have meaningful public health benefits, but these do not appear to reduce the cost of nutrient-adequate diets. This indicates that single nutrient fortification of starchy staples and oils or inefficiently high fortification levels for nutrients with ULs may cause the optimization model to include more expensive foods that allow the modeled diet to remain within all nutrient constraints. Fortification of multiple food vehicles with multiple nutrients at intermediate fortification levels is likely to avoid this problem as the role of any single food or nutrient is reduced.

The modeled diet cost reductions represent potential improvements in economic access to nutrient-adequate diets across a range of food group specifications. Compared to previous studies based on the basic CoNA analysis[17,19], which estimates the absolute lower-bound cost of achieving nutrient adequacy, this analysis introduces expanded scenarios that incorporate food group-level constraints. In particular, the CoNA-SUA scenario integrates large-scale consumption patterns derived from national Supply and Utilization Accounts, addressing critiques of the basic CoNA model's unrealistic assumptions by aligning diet modeling more closely with actual food supply distributions and cultural preferences. This scenario produces cost estimates that are higher than each of the other two scenarios, but also exhibits larger cost reduction percentages with LSFF implementation. The elevated cost likely reflects energy shares for comparatively expensive food groups for which consumption often overshoots food-based dietary guidelines, e.g., animal-source foods[25]. The CoNA-SUA scenario also produces cost estimates that approach the global average Cost of a Healthy Diet based on food-based dietary guidelines.[20] These findings suggest that implementation of LSFF policies could improve access to nutrient-adequate diets even under locally preferred food group consumption patterns. Moreover, our analysis of the changes in item inclusion dynamics further shows that when food group constraints are included, cost reductions occur without compromising overall diet quality, as fortified items generally replace their unfortified counterparts rather than altering the balance of the diet. Finally, results show that cost reduction effects are consistently greater for nutritionally vulnerable groups, namely women of reproductive age, adolescent boys, and older men.

This analysis has several limitations. First, the ICP food price data are nationally representative average prices, which ensures consistency and robustness but may obscure important local-level variation in food availability and affordability. Similarly, LSFF implementation may not reduce diet costs for all consumers within a given country, as some households, such as those that primarily purchase food at informal markets, may not have retail access to items subject to fortification. This analysis also does not account for all costs of LSFF implementation, such as costs to industry for equipment, labor, training, management, and administration, or costs to government related to monitoring and enforcement efforts. These costs are estimated in Friesen et al. (2026)[7] but are not capable of item-level mapping for diet cost modeling and are thus excluded from this analysis.

Second, this analysis was strictly limited to modeling of diet costs, and did not allow for comparisons with reported diets or food spending patterns. As such, a complete accounting of how modeled diets and observed behaviors either align or differ was not included. While the modeled diets are nutritionally adequate, and while the CoNA-SUA scenario was designed to approximate national dietary patterns, modeled diets may not fully align with national food-based dietary guidelines in each country. Future research could explore how incorporating both nutrient requirements and recommended consumption patterns into diet models may affect the estimated economic cost, particularly in scenarios that reflect healthy, diverse diets. Additional work could also investigate whether underlying food price patterns or particular aspects of LSFF policies explain the wide variation in country-level cost reductions reported here. Future work should also assess how cost reductions due to LSFF might improve the affordability of nutrient-adequate diets by comparing modeled diet costs with household purchasing power and the cost of basic needs.

These modeling findings show that implementation of LSFF generally produces only moderate reductions in the cost of nutrient-adequate diets and may in some cases have no cost effect or modestly increase diet costs. The extent of cost reductions varies widely across countries, fortification program design and extent, and population subgroups. The greatest potential reduction is observed where fortification levels are intermediate rather than high, baseline diet costs are relatively high, and modeled diets reflect national average consumption patterns, as well as among vulnerable populations with higher nutrient needs. These results underscore that LSFF alone cannot make diets affordable, but that carefully designed, context-specific programs can contribute meaningfully to improving access to adequate nutrient intake. Estimates of the extent of local nutrient deficiencies are also necessary for more sophisticated quantification of the benefits of LSFF, as improved health outcomes may outweigh the potential cost increases associated with high levels of fortification in places where nutrient deficiencies area extensive. Crucially, this analysis did not attempt to account for the full range of health and social benefits from LSFF, which are substantial[7], but focused exclusively on retail food costs as a measure of food security and access to nutritious diets. Accordingly, this work should not be taken to suggest that improved implementation of existing LSFF policies would not result in benefits to human health and wellbeing.

Experts should consider LSFF to be one element of a broader nutrition strategy, which requires the involvement of policymakers in government offices of health, agriculture, economic development, and other areas related to the regulation of the food industry. Fortification must also accompany other policy interventions to either improve household purchasing power or reduce the cost of nutritious foods in cases where prices are unusually high. Fortification policies should also be designed with regard for optimal fortification amounts, both to achieve local health goals and to avoid unintended excess intake of any given nutrient. Although the modeled diets do not match real-world consumption, this analysis suggests that well-designed LSFF programs may improve the affordability of nutrient-adequate diets. It also highlights the complementary role of fortification alongside broader food system or market-based efforts to expand access to healthy diets that fully align with food-based dietary guidelines.

## *Study information*

**Patient and public involvement**
This modeling study did not incorporate any patient or public involvement.

**Competing interests:** None declared.

**Ethics approval:** No human participants were involved in the generation of the data used and ethics approval was not required for the analysis of this data.

Supplemental Information for

# Impacts of large-scale food fortification on the cost of nutrient-adequate diets: a modelling study in 89 countries

Leah Costlow*[1,2], Yan Bai[3,4], Katherine P. Adams[5], Ty Beal[6], Kathryn G. Dewey[5], Christopher M. Free[6,7], Valerie M. Friesen[8], Mduduzi N. N. Mbuya[9], Stella Nordhagen[9], Florencia C. Vasta[9], William A. Masters[10,11]

**Table A1. Nutrients included in all least-cost diet calculations.**

| Panel A. Energy and macronutrients | |
|---|---|
| Nutrient | Type of constraint |
| Energy | Target quantity |
| Protein | Range |
| Lipids | Range |
| Carbohydrate | Range |
| **Panel B. Micronutrients** | |
| Calcium | Range |
| Choline | Range |
| Copper | Lower bound (pregnancy and lactation)<br>Range (all other subgroups) |
| Folate | Lower bound |
| Iron | Range |
| Magnesium | Lower bound |
| Manganese | Range |
| Niacin | Lower bound |
| Phosphorus | Range |
| Retinol | Upper bound |
| Riboflavin | Lower bound |
| Selenium | Range |
| Sodium | Upper bound |
| Thiamin | Lower bound |
| Vitamin A | Lower bound |
| Vitamin B5 | Lower bound |
| Vitamin B6 | Range |
| Vitamin B12 | Lower bound |
| Vitamin C | Range |
| Vitamin E | Range |
| Zinc | Range |

**Table A2. Sample matching of retail items to fortification vehicles**

| Country | Item | Bread (Y/N) | Fortifiable (Y/N) | Vehicle name |
|---|---|---|---|---|
| Kazakhstan | Rice (1) | No | No | N/A |
| Kazakhstan | Rice (2) | No | No | N/A |
| Kazakhstan | Rice (3) | No | No | N/A |
| Kazakhstan | Rice (4) | No | No | N/A |
| Kazakhstan | Wheat flour (1) | No | Yes | Wheat flour |
| Kazakhstan | Wheat flour (2) | No | Yes | Wheat flour |
| Kazakhstan | Wheat (1) | No | No | N/A |
| Kazakhstan | Bread (1) | Yes | Yes | Wheat flour |
| Kazakhstan | Bread (2) | Yes | Yes | Wheat flour |
| Kazakhstan | Bread (3) | Yes | Yes | Wheat flour |
| Kazakhstan | Bread (4) | Yes | Yes | Wheat flour |
| Zimbabwe | Rice (1) | No | No | N/A |
| Zimbabwe | Rice (2) | No | No | N/A |
| Zimbabwe | Rice (3) | No | No | N/A |
| Zimbabwe | Rice (4) | No | No | N/A |
| Zimbabwe | Wheat flour (1) | No | Yes | Wheat flour |
| Zimbabwe | Bread (1) | Yes | Yes | Wheat flour |
| Zimbabwe | Bread (2) | Yes | Yes | Wheat flour |
| Zimbabwe | Bread (3) | Yes | Yes | Wheat flour |
| Zimbabwe | Bread (4) | Yes | Yes | Wheat flour |

Note: Items names are generic versions to protect proprietary ICP item names and codes. Items are classified as fortifiable if matched to a food vehicle that is designated as subject to mandatory or voluntary fortification policies in the GFDx database.

**Table A3. Median, minimum, and maximum modeled diet costs with 0% and 90% LSFF implementation by model and region, including all 89 countries in analytical dataset.**

| Region | Cost (2021 PPP) | CoNA 0% LSFF | CoNA 90% LSFF | CoNA+SS&FV 0% LSFF | CoNA+SS&FV 90% LSFF | CoNA+SUA 0% LSFF | CoNA+SUA 90% LSFF |
|---|---|---|---|---|---|---|---|
| East Asia & Pacific | Median | 2.74 | 2.70 | 3.38 | 3.34 | 4.46 | 4.37 |
| | Minimum | 1.09 | 0.96 | 1.34 | 1.34 | 1.76 | 1.78 |
| | Maximum | 4.82 | 4.67 | 5.47 | 5.33 | 11.73 | 11.73 |
| Europe & Central Asia | Median | 2.29 | 2.27 | 2.65 | 2.63 | 4.00 | 3.79 |
| | Minimum | 0.90 | 0.72 | 1.00 | 0.80 | 1.54 | 1.40 |
| | Maximum | 3.88 | 3.89 | 4.35 | 4.35 | 8.77 | 8.77 |
| Latin America & Caribbean | Median | 3.33 | 3.13 | 4.08 | 3.76 | 4.82 | 4.17 |
| | Minimum | 1.69 | 1.40 | 1.87 | 1.69 | 1.84 | 1.66 |
| | Maximum | 6.11 | 5.65 | 8.17 | 7.72 | 9.54 | 8.37 |
| Middle East & North Africa | Median | 2.20 | 2.10 | 2.62 | 2.54 | 3.07 | 2.95 |
| | Minimum | 1.06 | 0.84 | 1.11 | 1.11 | 1.37 | 1.36 |
| | Maximum | 3.47 | 3.41 | 4.07 | 4.07 | 4.61 | 4.41 |
| North America | Median | 2.47 | 2.34 | 2.83 | 2.70 | 10.89 | 8.91 |
| | Minimum | 0.95 | 0.74 | 1.25 | 1.03 | 3.89 | 3.21 |
| | Maximum | 4.03 | 3.97 | 4.63 | 4.37 | 16.94 | 16.94 |
| South Asia | Median | 2.70 | 2.68 | 3.41 | 3.41 | 3.48 | 3.47 |
| | Minimum | 1.58 | 1.57 | 2.07 | 2.07 | 1.99 | 1.99 |
| | Maximum | 3.89 | 3.89 | 5.05 | 5.05 | 5.06 | 5.09 |
| Sub-Saharan Africa | Median | 2.51 | 2.51 | 2.70 | 2.69 | 3.13 | 3.08 |
| | Minimum | 1.28 | 1.28 | 1.31 | 1.31 | 1.51 | 1.51 |
| | Maximum | 4.51 | 4.51 | 4.58 | 4.58 | 5.62 | 5.47 |
| World | Median | 2.70 | 2.65 | 3.09 | 3.01 | 3.60 | 3.44 |
| | Minimum | 0.90 | 0.72 | 1.00 | 0.80 | 1.37 | 1.36 |
| | Maximum | 6.11 | 5.65 | 8.17 | 7.72 | 16.94 | 16.94 |

Note: Data shown are diet cost estimates calculated across 3 nutrient adequacy scenarios using retail food prices from 2021. World regions are from the World Bank Development Indicators. Diet costs are shown in 2021 international dollars (PPP). LSFF refers to large-scale food fortification. CoNA refers to diets meeting nutrient requirements; CoNA-SS&FV refers to diets that also meet basic healthy food group constraints for starchy staples, fruits, and vegetables; and CoNA-SUA refers to diets that align with national average consumption patterns at the food group level.

**Table A4. Region and income group classifications for countries included in fortification analysis**

| Country | ISO3 | Region | Income group |
|---|---|---|---|
| Antigua and Barbuda | ATG | Latin America & Caribbean | High income |
| Argentina | ARG | Latin America & Caribbean | Upper-middle income |
| Australia | AUS | East Asia & Pacific | High income |
| Bahrain | BHR | Middle East & North Africa | High income |
| Bangladesh | BGD | South Asia | Lower-middle income |
| Belize | BLZ | Latin America & Caribbean | Upper-middle income |
| Benin | BEN | Sub-Saharan Africa | Lower-middle income |
| Bolivia | BOL | Latin America & Caribbean | Lower-middle income |
| Brazil | BRA | Latin America & Caribbean | Upper-middle income |
| Burkina Faso | BFA | Sub-Saharan Africa | Lower income |
| Burundi | BDI | Sub-Saharan Africa | Lower income |
| Cabo Verde | CPV | Sub-Saharan Africa | Lower-middle income |
| Cameroon | CMR | Sub-Saharan Africa | Lower-middle income |
| Canada | CAN | North America | High income |
| Chad | TCD | Sub-Saharan Africa | Lower income |
| Chile | CHL | Latin America & Caribbean | High income |
| China | CHN | East Asia & Pacific | Upper-middle income |
| Colombia | COL | Latin America & Caribbean | Upper-middle income |
| Congo | COG | Sub-Saharan Africa | Lower-middle income |
| Costa Rica | CRI | Latin America & Caribbean | Upper-middle income |
| Cote d'Ivoire | CIV | Sub-Saharan Africa | Lower-middle income |
| Djibouti | DJI | Middle East & North Africa | Lower-middle income |
| Dominica | DMA | Latin America & Caribbean | Upper-middle income |
| Dominican Republic | DOM | Latin America & Caribbean | Upper-middle income |
| Ecuador | ECU | Latin America & Caribbean | Upper-middle income |
| El Salvador | SLV | Latin America & Caribbean | Lower-middle income |
| Eswatini | SWZ | Sub-Saharan Africa | Lower-middle income |
| Ethiopia | ETH | Sub-Saharan Africa | Lower income |
| Fiji | FJI | East Asia & Pacific | Upper-middle income |
| Gambia | GMB | Sub-Saharan Africa | Lower income |
| Ghana | GHA | Sub-Saharan Africa | Lower-middle income |
| Grenada | GRD | Latin America & Caribbean | Upper-middle income |
| Guinea | GIN | Sub-Saharan Africa | Lower income |
| Guyana | GUY | Latin America & Caribbean | Upper-middle income |
| Honduras | HND | Latin America & Caribbean | Lower-middle income |
| India | IND | South Asia | Lower-middle income |
| Indonesia | IDN | East Asia & Pacific | Lower-middle income |
| Iraq | IRQ | Middle East & North Africa | Upper-middle income |
| Jamaica | JAM | Latin America & Caribbean | Upper-middle income |
| Jordan | JOR | Middle East & North Africa | Upper-middle income |
| Kazakhstan | KAZ | Europe & Central Asia | Upper-middle income |
| Kenya | KEN | Sub-Saharan Africa | Lower-middle income |
| Kuwait | KWT | Middle East & North Africa | High income |
| Liberia | LBR | Sub-Saharan Africa | Lower income |
| Malawi | MWI | Sub-Saharan Africa | Lower income |
| Malaysia | MYS | East Asia & Pacific | Upper-middle income |
| Mali | MLI | Sub-Saharan Africa | Lower income |
| Mexico | MEX | Latin America & Caribbean | Upper-middle income |
| Moldova | MDA | Europe & Central Asia | Upper-middle income |
| Mongolia | MNG | East Asia & Pacific | Lower-middle income |
| Morocco | MAR | Middle East & North Africa | Lower-middle income |
| Mozambique | MOZ | Sub-Saharan Africa | Lower income |
| Nepal | NPL | South Asia | Lower-middle income |
| Netherlands | NLD | Europe & Central Asia | High income |
| New Zealand | NZL | East Asia & Pacific | High income |
| Nicaragua | NIC | Latin America & Caribbean | Lower-middle income |
| Niger | NER | Sub-Saharan Africa | Lower income |
| Nigeria | NGA | Sub-Saharan Africa | Lower-middle income |
| Oman | OMN | Middle East & North Africa | High income |
| Pakistan | PAK | South Asia | Lower-middle income |
| Panama | PAN | Latin America & Caribbean | High income |
| Paraguay | PRY | Latin America & Caribbean | Upper-middle income |
| Peru | PER | Latin America & Caribbean | Upper-middle income |
| Philippines | PHL | East Asia & Pacific | Lower-middle income |
| Qatar | QAT | Middle East & North Africa | High income |
| Rwanda | RWA | Sub-Saharan Africa | Lower income |
| Saint Kitts and Nevis | KNA | Latin America & Caribbean | High income |
| Saint Lucia | LCA | Latin America & Caribbean | Upper-middle income |
| Saint Vincent and the Grenadines | VCT | Latin America & Caribbean | Upper-middle income |
| Saudi Arabia | SAU | Middle East & North Africa | High income |
| Senegal | SEN | Sub-Saharan Africa | Lower-middle income |
| Sierra Leone | SLE | Sub-Saharan Africa | Lower income |
| South Africa | ZAF | Sub-Saharan Africa | Upper-middle income |
| Sri Lanka | LKA | South Asia | Lower-middle income |

| Sudan | SDN | Sub-Saharan Africa | Lower income |
|---|---|---|---|
| Suriname | SUR | Latin America & Caribbean | Upper-middle income |
| Tanzania | TZA | Sub-Saharan Africa | Lower-middle income |
| Thailand | THA | East Asia & Pacific | Upper-middle income |
| Togo | TGO | Sub-Saharan Africa | Lower income |
| Trinidad and Tobago | TTO | Latin America & Caribbean | High income |
| Uganda | UGA | Sub-Saharan Africa | Lower income |
| United Arab Emirates | ARE | Middle East & North Africa | High income |
| United Kingdom | GBR | Europe & Central Asia | High income |
| United States of America | USA | North America | High income |
| Uruguay | URY | Latin America & Caribbean | High income |
| Uzbekistan | UZB | Europe & Central Asia | Lower-middle income |
| Viet Nam | VNM | East Asia & Pacific | Lower-middle income |
| Zambia | ZMB | Sub-Saharan Africa | Lower income |
| Zimbabwe | ZWE | Sub-Saharan Africa | Lower-middle income |

**Table A5. Data availability and fortification coverage (n = country).**

**Panel A. Data coverage**

| | |
|---|---|
| International Comparisons Program (ICP) 2021[20] | 173 |
| Global Fortification Data Exchange (GFDx)[21] | 154 |
| FAO Supply Utilization Accounts (SUA)[19] | 190 |
| Final analytical sample | 89 |

**Panel B. Fortification policies by nutrient and food vehicle**

| **Nutrient** | **Food Vehicle** | | | | |
|---|---|---|---|---|---|
| | Wheat flour | Maize flour | Oil | Rice | Any vehicle |
| Iron | 82 | 18 | 0 | 10 | 82 |
| Zinc | 30 | 11 | 0 | 6 | 34 |
| Calcium | 23 | 2 | 0 | 2 | 23 |
| Vitamin B12 | 21 | 10 | 0 | 6 | 25 |
| Folate | 71 | 17 | 0 | 9 | 71 |
| Niacin | 59 | 16 | 0 | 10 | 61 |
| Riboflavin | 59 | 15 | 0 | 4 | 61 |
| Thiamin | 61 | 16 | 0 | 11 | 63 |
| Vitamin B6 | 16 | 9 | 0 | 5 | 20 |
| Vitamin A | 21 | 10 | 36 | 3 | 43 |
| Vitamin E | 0 | 0 | 2 | 1 | 3 |
| Selenium | 1 | 0 | 0 | 1 | 2 |
| Any nutrient | 86 | 18 | 37 | 12 | 89 |

Note: Fortification policies represent unique country, nutrient, and food vehicle combinations, including both mandatory and voluntary standards as documented in GFDx.

**Table A6. Diet costs with alternative rates of LSFF implementation and premix cost pass-through**

A. CoNA

| | Base | 90%LSFF + 50% PT | 50% LSFF + 100% PT | 90% LSFF + 100% PT |
|---|---|---|---|---|
| East Asia & Pacific | 2.74 (0.93) | 2.72 (1.02) | 2.73 (1.02) | 2.72 (1.02) |
| Europe & Central Asia | 2.24 (0.70) | 2.24 (0.79) | 2.23 (0.75) | 2.24 (0.79) |
| Latin America & Caribbean | 3.32 (0.87) | 3.10 (0.89) | 3.17 (0.87) | 3.10 (0.89) |
| Middle East & North Africa | 2.20 (0.71) | 2.09 (0.81) | 2.14 (0.74) | 2.09 (0.80) |
| North America | 2.30 (1.47) | 2.12 (1.65) | 2.13 (1.65) | 2.13 (1.65) |
| South Asia | 2.72 (0.65) | 2.70 (0.68) | 2.68 (0.68) | 2.70 (0.68) |
| Sub-Saharan Africa | 2.48 (0.72) | 2.48 (0.72) | 2.48 (0.72) | 2.48 (0.72) |

B. CoNA-SS&FV

| | Base | 90%LSFF + 50% PT | 50% LSFF + 100% PT | 90% LSFF + 100% PT |
|---|---|---|---|---|
| East Asia & Pacific | 3.41 (1.04) | 3.36 (1.13) | 3.38 (1.11) | 3.36 (1.13) |
| Europe & Central Asia | 2.63 (1.10) | 2.63 (1.13) | 2.62 (1.13) | 2.63 (1.13) |
| Latin America & Caribbean | 4.08 (1.12) | 3.74 (1.10) | 3.82 (1.07) | 3.74 (1.10) |
| Middle East & North Africa | 2.63 (0.80) | 2.52 (0.81) | 2.56 (0.76) | 2.53 (0.81) |
| North America | 2.70 (1.32) | 2.49 (1.52) | 2.49 (1.50) | 2.49 (1.52) |
| South Asia | 3.36 (1.08) | 3.39 (1.06) | 3.35 (1.06) | 3.39 (1.06) |
| Sub-Saharan Africa | 2.69 (0.80) | 2.69 (0.81) | 2.69 (0.80) | 2.69 (0.81) |

C. CoNA-SUA

| | Base | 90%LSFF + 50% PT | 50% LSFF + 100% PT | 90% LSFF + 100% PT |
|---|---|---|---|---|
| East Asia & Pacific | 4.03 (1.41) | 3.92 (1.46) | 3.92 (1.47) | 3.92 (1.46) |
| Europe & Central Asia | 4.06 (1.49) | 3.84 (1.67) | 3.93 (1.57) | 3.84 (1.67) |
| Latin America & Caribbean | 4.73 (1.69) | 4.09 (1.35) | 4.34 (1.53) | 4.09 (1.35) |
| Middle East & North Africa | 3.04 (1.12) | 2.93 (0.96) | 2.99 (1.02) | 2.93 (0.96) |
| North America | 10.84 (6.51) | 6.92 (7.51) | 8.14 (6.87) | 6.92 (7.51) |
| South Asia | 3.45 (0.87) | 3.45 (0.88) | 3.45 (0.87) | 3.45 (0.88) |
| Sub-Saharan Africa | 3.09 (0.80) | 3.06 (0.83) | 3.06 (0.83) | 3.06 (0.83) |

Note: Data shown are median diet cost estimates (interquartile ranges) in 2021 international dollars (PPP). Base = No fortification; Fort 1 = sensitivity analysis with 90% LSFF implementation and 50% pass-through of premix costs; Fort 2 = sensitivity analysis with 50% LSFF implementation and 100% pass-through of premix costs; and Fort 3 = main fortification analysis with 90% LSFF implementation and 100% pass-through of premix costs. CoNA refers to diets meeting nutrient requirements; CoNA-SS&FV refers to diets that also meet basic healthy food group constraints for starchy staples, fruits, and vegetables; and CoNA-SUA refers to diets that align with national average consumption patterns at the food group level.

**Table A7. Modeled diet cost scenarios with no feasible solution by country and subgroup**

| Country | Subgroup | Scenario |
|---|---|---|
| Australia | Male (11 to 14) | CoNA-SUA (no LSFF) |
| Australia | Male (11 to 14) | CoNA-SUA (with LSFF) |
| Bangladesh | Male (71 to 79) | CoNA-SUA (no LSFF) |
| Bangladesh | Male (71 to 79) | CoNA-SUA (with LSFF) |
| Canada | Female (11 to 14) | CoNA-SUA (no LSFF) |
| Canada | Female (11 to 14) | CoNA-SUA (with LSFF) |
| Canada | Female (15 to 17) | CoNA-SUA (no LSFF) |
| Canada | Female (15 to 17) | CoNA-SUA (with LSFF) |
| Canada | Female (18 to 24) | CoNA-SUA (no LSFF) |
| Canada | Female (18 to 24) | CoNA-SUA (with LSFF) |
| Canada | Female (25 to 50) | CoNA-SUA (no LSFF) |
| Canada | Female (25 to 50) | CoNA-SUA (with LSFF) |
| Canada | Female (51 to 70) | CoNA-SUA (no LSFF) |
| Canada | Female (51 to 70) | CoNA-SUA (with LSFF) |
| Canada | Female (7 to 10) | CoNA-SUA (no LSFF) |
| Canada | Female (7 to 10) | CoNA-SUA (with LSFF) |
| Canada | Female (71 to 79) | CoNA-SUA (no LSFF) |
| Canada | Female (71 to 79) | CoNA-SUA (with LSFF) |
| Canada | Male (11 to 14) | CoNA-SUA (no LSFF) |
| Canada | Male (11 to 14) | CoNA-SUA (with LSFF) |
| Canada | Male (15 to 17) | CoNA-SUA (no LSFF) |
| Canada | Male (15 to 17) | CoNA-SUA (with LSFF) |
| El Salvador | Female (11 to 14) | CoNA-SUA (no LSFF) |
| El Salvador | Female (11 to 14) | CoNA-SUA (with LSFF) |
| El Salvador | Female (15 to 17) | CoNA-SUA (no LSFF) |
| El Salvador | Female (15 to 17) | CoNA-SUA (with LSFF) |
| El Salvador | Female (18 to 24) | CoNA-SUA (no LSFF) |
| El Salvador | Female (18 to 24) | CoNA-SUA (with LSFF) |
| El Salvador | Female (25 to 50) | CoNA-SUA (no LSFF) |
| El Salvador | Female (25 to 50) | CoNA-SUA (with LSFF) |
| El Salvador | Female (4 to 6) | CoNA-SUA (no LSFF) |
| El Salvador | Female (4 to 6) | CoNA-SUA (with LSFF) |
| El Salvador | Female (51 to 70) | CoNA-SUA (no LSFF) |
| El Salvador | Female (51 to 70) | CoNA-SUA (with LSFF) |
| El Salvador | Female (7 to 10) | CoNA-SUA (no LSFF) |
| El Salvador | Female (7 to 10) | CoNA-SUA (with LSFF) |
| El Salvador | Female (71 to 79) | CoNA-SUA (no LSFF) |
| El Salvador | Female (71 to 79) | CoNA-SUA (with LSFF) |
| El Salvador | Lactating (19 to 30) | CoNA-SUA (no LSFF) |
| El Salvador | Lactating (19 to 30) | CoNA-SUA (with LSFF) |
| El Salvador | Lactating (31 to 50) | CoNA-SUA (no LSFF) |
| El Salvador | Lactating (31 to 50) | CoNA-SUA (with LSFF) |
| El Salvador | Lactating (18 and under) | CoNA-SUA (no LSFF) |
| El Salvador | Lactating (18 and under) | CoNA-SUA (with LSFF) |
| El Salvador | Male (11 to 14) | CoNA-SUA (no LSFF) |
| El Salvador | Male (11 to 14) | CoNA-SUA (with LSFF) |
| El Salvador | Male (15 to 17) | CoNA-SUA (no LSFF) |
| El Salvador | Male (15 to 17) | CoNA-SUA (with LSFF) |
| El Salvador | Male (18 to 24) | CoNA-SUA (no LSFF) |
| El Salvador | Male (18 to 24) | CoNA-SUA (with LSFF) |
| El Salvador | Male (25 to 50) | CoNA-SUA (no LSFF) |
| El Salvador | Male (25 to 50) | CoNA-SUA (with LSFF) |
| El Salvador | Male (4 to 6) | CoNA-SUA (no LSFF) |
| El Salvador | Male (4 to 6) | CoNA-SUA (with LSFF) |
| El Salvador | Male (51 to 70) | CoNA-SUA (no LSFF) |
| El Salvador | Male (51 to 70) | CoNA-SUA (with LSFF) |
| El Salvador | Male (7 to 10) | CoNA-SUA (no LSFF) |
| El Salvador | Male (7 to 10) | CoNA-SUA (with LSFF) |
| El Salvador | Male (71 to 79) | CoNA-SUA (no LSFF) |
| El Salvador | Male (71 to 79) | CoNA-SUA (with LSFF) |
| El Salvador | Pregnant (19 to 30) | CoNA-SUA (no LSFF) |
| El Salvador | Pregnant (19 to 30) | CoNA-SUA (with LSFF) |
| El Salvador | Pregnant (31 to 50) | CoNA-SUA (no LSFF) |
| El Salvador | Pregnant (31 to 50) | CoNA-SUA (with LSFF) |
| El Salvador | Pregnant (18 and under) | CoNA-SUA (no LSFF) |
| El Salvador | Pregnant (18 and under) | CoNA-SUA (with LSFF) |
| United States of America | Female (11 to 14) | CoNA-SUA (no LSFF) |
| United States of America | Female (25 to 50) | CoNA-SUA (no LSFF) |
| United States of America | Female (7 to 10) | CoNA-SUA (no LSFF) |
| United States of America | Male (11 to 14) | CoNA-SUA (no LSFF) |

Note: CoNA-SUA refers to diets that align with national average consumption patterns at the food group level. LSFF refers to large-scale food fortification.

**Figure A1. Food group inclusion without and with fortification relative to Healthy Diet Basket reference intakes**

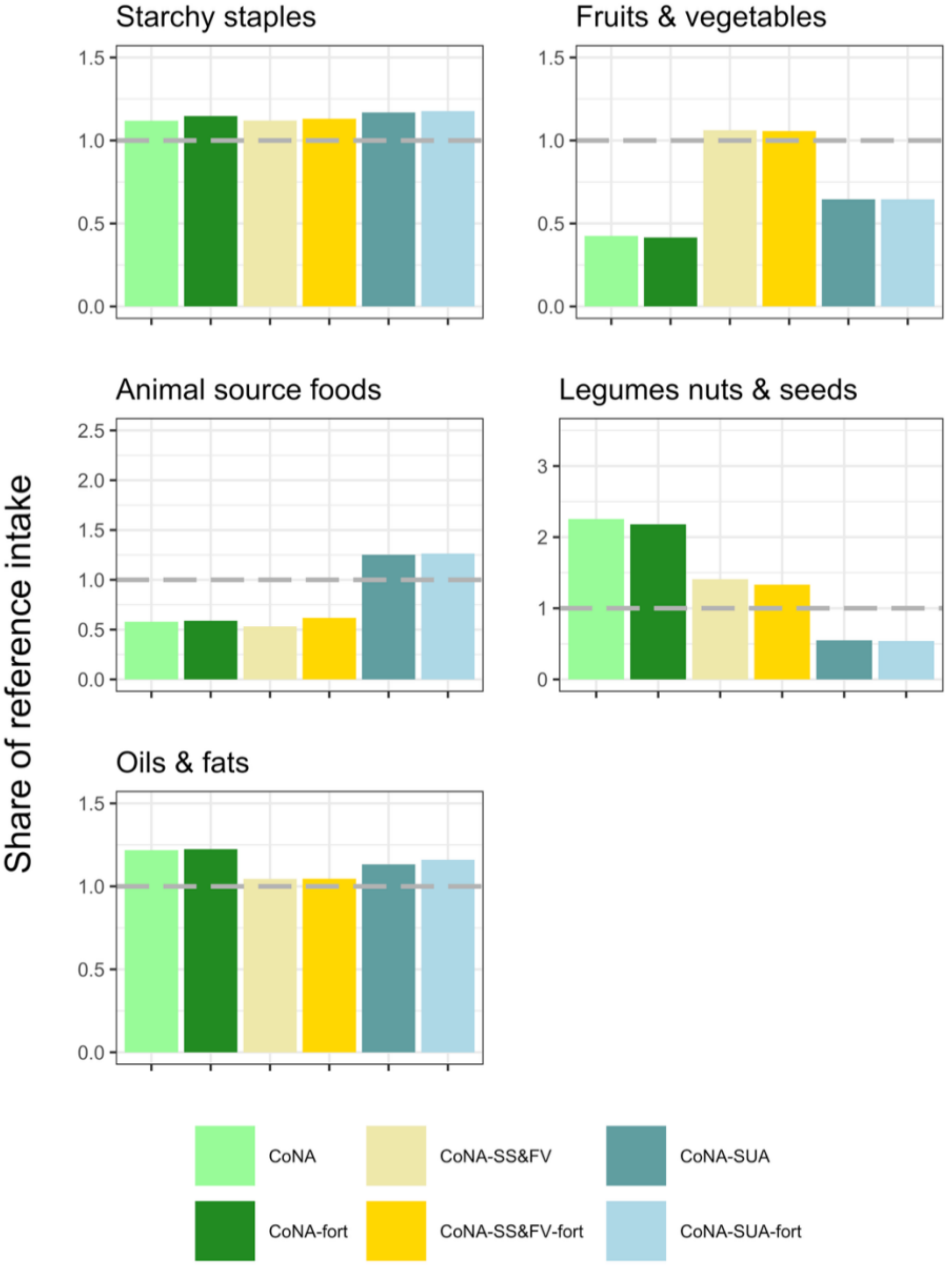

**Note:** Data shown are average kilocalories per day included in modeled diets by food group, expressed as the ratio of reference intakes from the Healthy Diet Basket[13]. Reference intakes are adjusted proportionally for each sex-age subgroup's total energy needs. Dashed horizontal lines indicate where food group consumption is equal to reference intakes. Ratios are calculated at the model-subgroup level and then converted to the overall averages shown here. CoNA refers to diets meeting nutrient requirements; CoNA-SS&FV refers to diets that also meet basic healthy food group constraints for starchy staples, fruits, and vegetables; and CoNA-SUA refers to diets that align with national average consumption patterns at the food group level. The suffix "-fort" indicates that 90% LSFF implementation has been included in the diet cost modeling scenario.

**Figure A2. Variation in country-level diet cost reductions with improved scenario by scenario and number of fortification policies.**

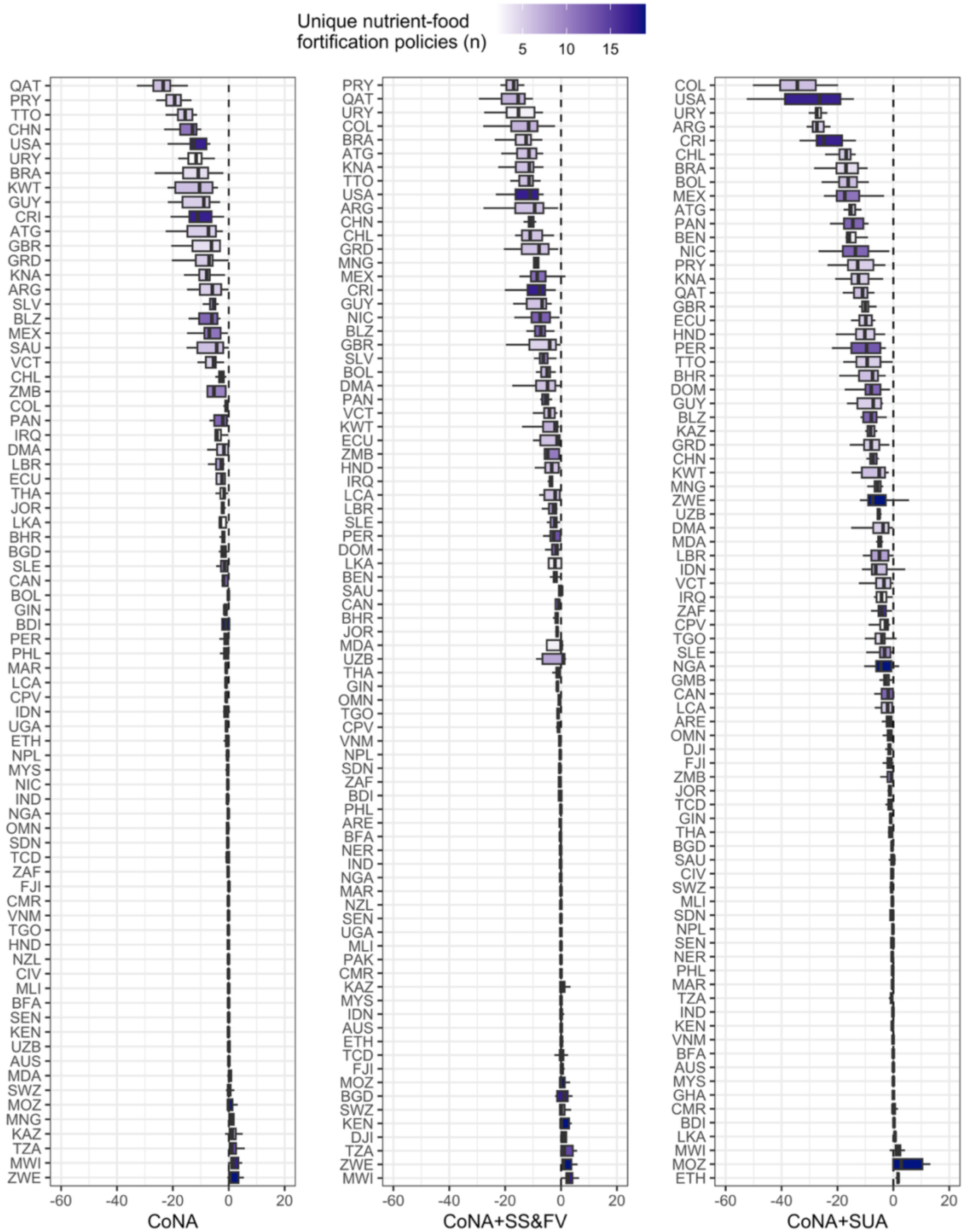

**Note:** Data shown are changes in diet costs from countries with non-zero change in diet costs under LSFF implementation. Boxplots indicate the distribution of diet cost changes within each country across 22 sex-age subgroups. Negative values indicate reduced costs with LSFF while positive values indicate increased costs with LSFF. Shading within each box indicates the number of unique fortification policies documented in each country, e.g. the total number of policies per food vehicle and nutrient combination. Y-axis labels are ISO3 country codes. CoNA refers to diets meeting nutrient requirements; CoNA-SS&FV refers to diets that also meet basic healthy food group constraints for starchy staples, fruits, and vegetables; and CoNA-SUA refers to diets that align with national average consumption patterns at the food group level. Outliers are omitted from all boxplots using the Tukey standard (1.5 x

IQR).

**Figure A3. Examples of food item inclusion with and without fortification under CoNA+SUA scenario for girls aged 4y to 6y.**

**Ethiopia**

*(wheat flour and oil subject to fortification)*

No fortification
90% fortification

0.0 0.5 1.0 1.5 2.0 2.5

Bread
Bread (fortified)
Cabbage
Cassava
Cassava leaves
Collard
Eggs
Maize
Organ meat
Palm oil
Pasta
Peanuts
Rice
Shrimp
Spotted beans

**Togo**

*(wheat flour and oil subject to fortification)*

No fortification
90% fortification

0.0 0.5 1.0 1.5 2.0 2.5

Bread
Bread (fortified)
Cassava
Cassava leaves
Eggs
Maize
Oil (fortified)
Organ meat
Palm oil
Pasta
Peanuts
Rice
Sardines
Shrimp
Spotted beans
Sweet potatoes
White beans

**Malawi**

*(wheat flour, maize flour, and oil subject to fortification)*

No fortification
90% fortification

0.0 0.5 1.0 1.5 2.0 2.5

Cost (2021 PPP per day)

Bean leaves
Cabbage
Cassava
Cassava leaves
Coconuts
Eggs
Maize
Oil (fortified)
Organ meat
Peanut oil
Peanuts
Rice
Sardines

**Note:** Data shown are cost per day in 2021 international dollars (PPP) to purchase items included in diets meeting CoNA+SUA with and without fortification for girls aged 4 to 6. Costs are shown for three selected countries where fortification increases average diet costs. CoNA-SUA refers to diets that align with national average consumption patterns at the food group level.

**Figure A4. Fortification premix costs as a share of retail prices by food vehicle**

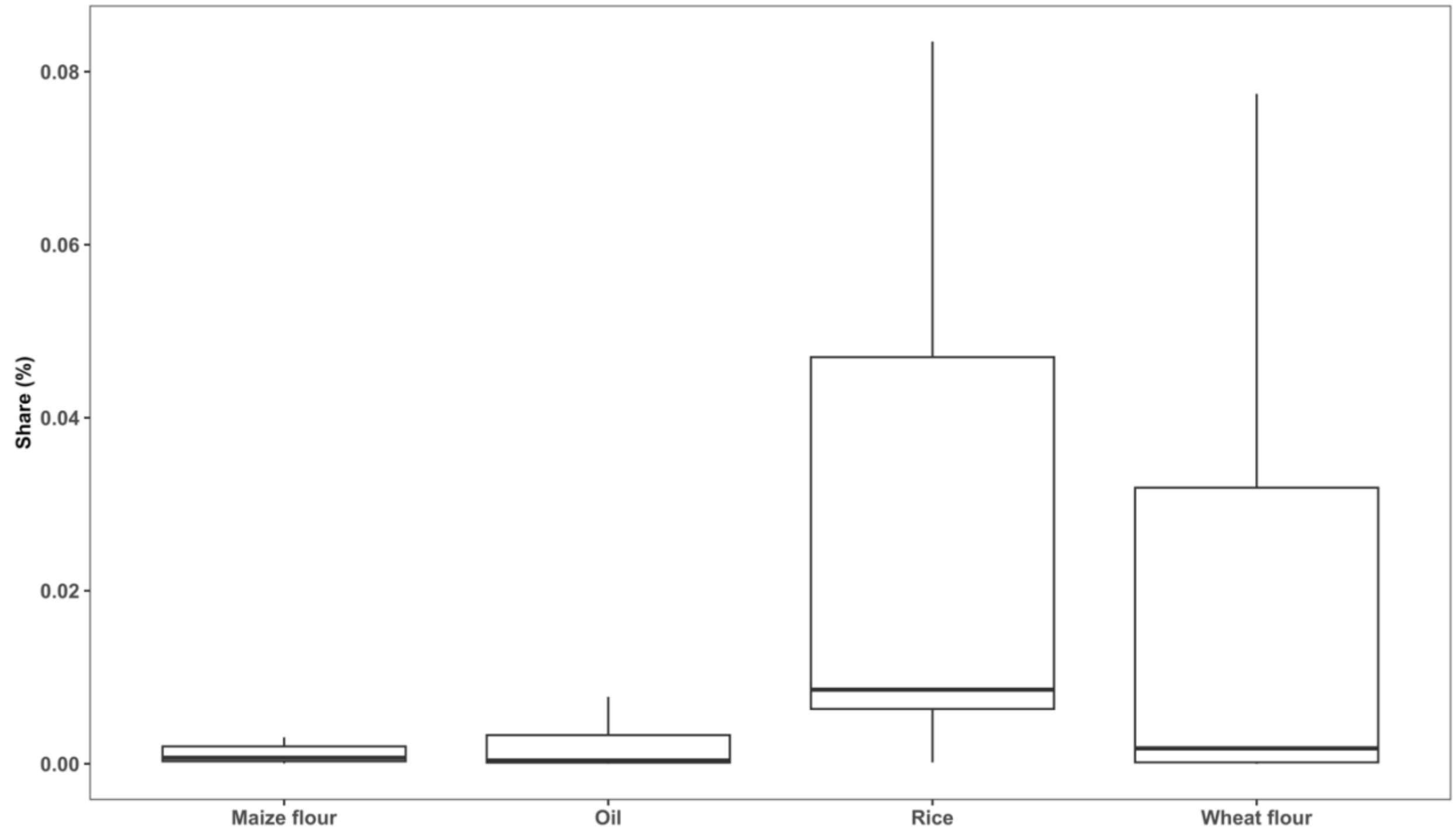

**Note:** Data shown are premix costs as estimated in Friesen et al. (2026) as a percentage of retail food prices for foods matched to fortification vehicles at the country level. Prices and premix costs are each specified in 2021 international dollars (PPP) per kilogram of fortified food vehicle. Outliers are omitted from all boxplots using the Tukey standard (1.5 x IQR).

**Figure A5. Percentage change in diet costs by country income group (all countries and population subgroups)**

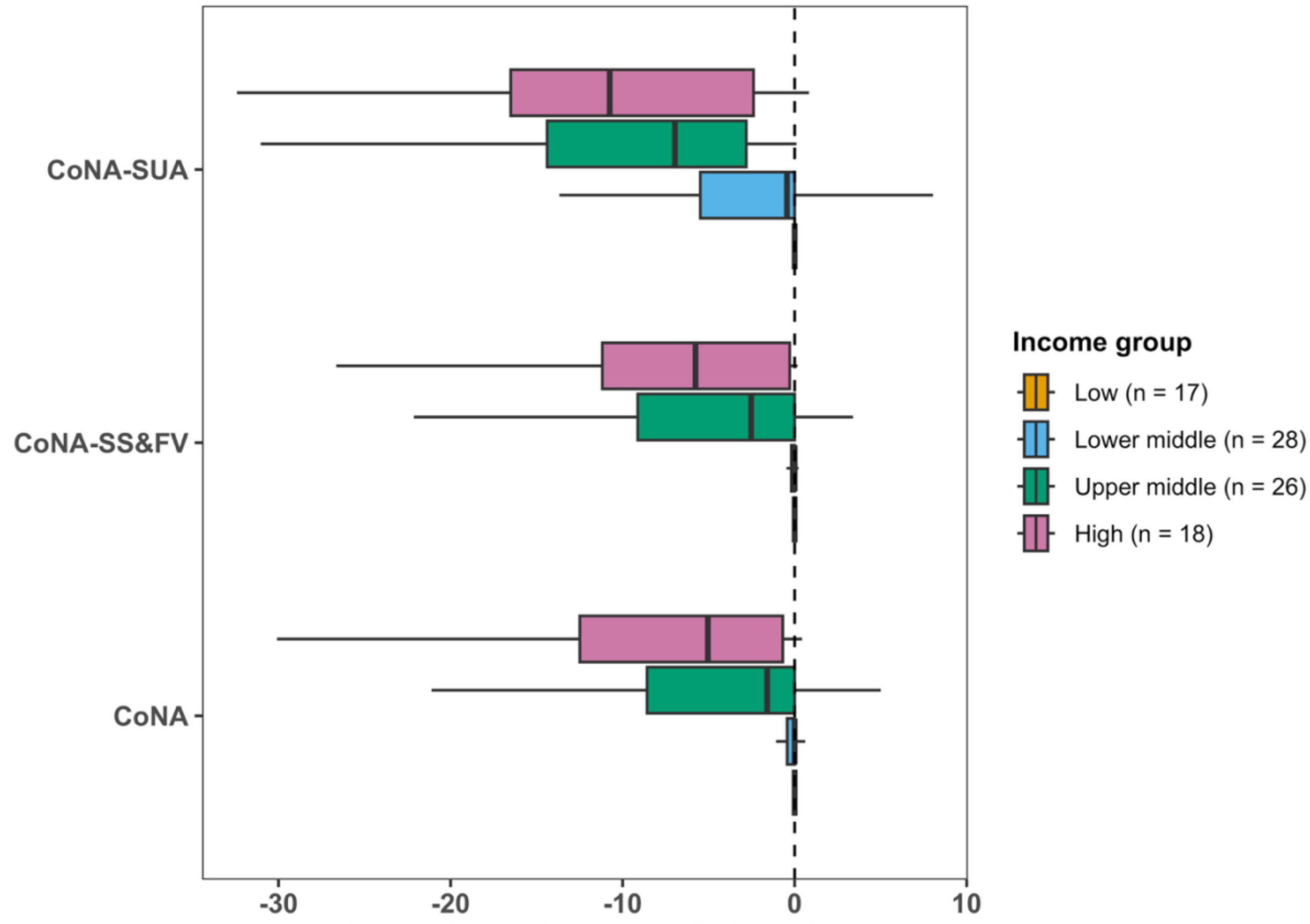

**Note:** Data shown are changes in diet costs under LSFF implementation, with all 89 sample countries included in the sample. Boxplots indicate the distribution of diet cost changes within each country across 22 sex-age subgroups. Negative values indicate reduced costs with LSFF while positive values indicate increased costs with LSFF. CoNA refers to diets meeting nutrient requirements; CoNA-SS&FV refers to diets that also meet basic healthy food group constraints for starchy staples, fruits, and vegetables; and CoNA-SUA refers to diets that align with national average consumption patterns at the food group level. Outliers are omitted from all boxplots using the Tukey standard (1.5 x IQR).